\documentclass[a4paper,manyauthors,nocleardouble,COMPASS]{cernphprep}
\usepackage{lineno}
\usepackage{subfigure}
\usepackage{ulem}
\usepackage{color}	
\definecolor{dimgray}{rgb}{0.31, 0.31, 0.21}
\definecolor{gray}{rgb}{0.41, 0.41, 0.41}
\usepackage{multirow}
\usepackage{makecell}
\usepackage{textcomp}
\usepackage{slashbox,booktabs,amsmath}

\newcommand{\dd}{\mathop{}\mathopen{}\mathrm{d}}

\begin{document}

\begin{titlepage}

\PHnumber{2017--xxx}
\PHdate{\today}

\title{Transverse-momentum-dependent Multiplicities of Charged Hadrons in Muon-Deuteron Deep Inelastic Scattering}

\Collaboration{The COMPASS Collaboration}
\ShortAuthor{The COMPASS Collaboration}

%%%%%%%%%%%%
% 	Abstract		 %
\begin{abstract}

A semi-inclusive measurement of charged hadron multiplicities in deep inelastic muon scattering off an isoscalar target was performed using data collected by the COMPASS Collaboration at CERN. The following kinematic domain is covered by the data: photon virtuality $Q^{2}>1$~(GeV/$c$)$^2$, invariant mass of the hadronic system $W > 5$ GeV/$c^2$, Bjorken scaling variable in the range $0.003 < x < 0.4$, fraction of the virtual photon energy carried by the hadron in the range $0.2 < z < 0.8$, square of the hadron transverse momentum with respect to the virtual photon direction in the range 0.02~(GeV/$c)^2 < P_{\rm{hT}}^{2} < 3$~(GeV/$c$)$^2$.
 The multiplicities are presented as a function of $P_{\rm{hT}}^{2}$ in three-dimensional bins of $x$, $Q^2$, $z$ and compared to previous semi-inclusive measurements. We explore the small-$P_{\rm{hT}}^{2}$ region,  i.e. $P_{\rm{hT}}^{2} < 1$~(GeV/$c$)$^2$, where hadron transverse momenta are expected to arise from non-perturbative effects, and also the domain of larger $P_{\rm{hT}}^{2}$, where contributions from higher-order perturbative QCD are expected to dominate.
The multiplicities are fitted using a single-exponential function at small $P_{\rm{hT}}^{2}$ to study the dependence of the average transverse momentum $\langle P_{\rm{hT}}^{2}\rangle$ on $x$, $Q^2$ and $z$. The power-law behaviour of the multiplicities at large $P_{\rm{hT}}^{2}$ is investigated using various functional forms. 
The fits describe the data reasonably well over the full measured range.

\end{abstract}

\vspace*{60pt}

\vfill
\Submitted{(to be submitted to Phys. Rev. D)}

\end{titlepage}

{\pagestyle{empty} %%%%%%%%%%%%%%%%%%%%%%%%%%%%%%%%%%%%%%%%%%%%%%%%%%%%%%%%%%%%%%%
%
% 2017_auththorlist.tex   29.05.2017
%
%%%%%%%%%%%%%%%%%%%%%%%%%%%%%%%%%%%%%%%%%%%%%%%%%%%%%%%%%%%%%%%
\section*{The COMPASS Collaboration}
\label{app:collab}
\renewcommand\labelenumi{\textsuperscript{\theenumi}~}
\renewcommand\theenumi{\arabic{enumi}}
\begin{flushleft}
M.~Aghasyan\Irefn{triest_i},
%R.~Akhunzyanov\Irefn{dubna}, % phd
M.G.~Alexeev\Irefn{turin_u},
G.D.~Alexeev\Irefn{dubna}, %1
A.~Amoroso\Irefnn{turin_u}{turin_i},
V.~Andrieux\Irefnn{illinois}{saclay},
N.V.~Anfimov\Irefn{dubna}, %2
V.~Anosov\Irefn{dubna}, %3
A.~Antoshkin\Irefn{dubna}, % phd
K.~Augsten\Irefnn{dubna}{praguectu}, %prague
W.~Augustyniak\Irefn{warsaw}
% ...
\end{flushleft}
%%%%%%%%%%%%%%%%%%%%%%%%%%%%%%%%%%%%%%%%%%%%%%%%%%%%%%%%%%%%%%%%%%%%%%%%%%%%%%%%%%%%%%%%%%%%%%%%%%%%%%%%%%%%%%%%%%%%%%%
%
% institutes
%
%%%%%%%%%%%%%%%%%%%%%%%%%%%%%%%%%%%%%%%%%%%%%%%%%%%%%%%%%%%%%%%%%%%%%%%%%%%%%%%%%%%%%%%%%%%%%%%%%%%%%%%%%%%%%%%%%%%%%%%
\begin{Authlist}
\item \Idef{aveiro}{University of Aveiro, Dept.\ of Physics, 3810-193 Aveiro, Portugal}
\item \Idef{bochum}{Universit\"at Bochum, Institut f\"ur Experimentalphysik, 44780 Bochum, Germany\Arefs{l}\Aref{s}}
\item \Idef{bonniskp}{Universit\"at Bonn, Helmholtz-Institut f\"ur  Strahlen- und Kernphysik, 53115 Bonn, Germany\Arefs{l}}
\item \Idef{bonnpi}{Universit\"at Bonn, Physikalisches Institut, 53115 Bonn, Germany\Arefs{l}}
\item \Idef{brno}{Institute of Scientific Instruments, AS CR, 61264 Brno, Czech Republic\Arefs{m}}
\item \Idef{calcutta}{Matrivani Institute of Experimental Research \& Education, Calcutta-700 030, India\Arefs{n}}
\item \Idef{dubna}{Joint Institute for Nuclear Research, 141980 Dubna, Moscow region, Russia\Arefs{o}}
\item \Idef{freiburg}{Universit\"at Freiburg, Physikalisches Institut, 79104 Freiburg, Germany\Arefs{l}\Aref{s}}
\item \Idef{cern}{CERN, 1211 Geneva 23, Switzerland}
\item \Idef{liberec}{Technical University in Liberec, 46117 Liberec, Czech Republic\Arefs{m}}
\item \Idef{lisbon}{LIP, 1000-149 Lisbon, Portugal\Arefs{p}}
\item \Idef{mainz}{Universit\"at Mainz, Institut f\"ur Kernphysik, 55099 Mainz, Germany\Arefs{l}}
\item \Idef{miyazaki}{University of Miyazaki, Miyazaki 889-2192, Japan\Arefs{q}}
\item \Idef{moscowlpi}{Lebedev Physical Institute, 119991 Moscow, Russia}
\item \Idef{munichtu}{Technische Universit\"at M\"unchen, Physik Dept., 85748 Garching, Germany\Arefs{l}\Aref{r}}
\item \Idef{nagoya}{Nagoya University, 464 Nagoya, Japan\Arefs{q}}
\item \Idef{praguecu}{Charles University in Prague, Faculty of Mathematics and Physics, 18000 Prague, Czech Republic\Arefs{m}}
\item \Idef{praguectu}{Czech Technical University in Prague, 16636 Prague, Czech Republic\Arefs{m}}
\item \Idef{protvino}{State Scientific Center Institute for High Energy Physics of National Research Center `Kurchatov Institute', 142281 Protvino, Russia}
\item \Idef{saclay}{IRFU, CEA, Universit\'e Paris-Saclay, 91191 Gif-sur-Yvette, France\Arefs{s}}
\item \Idef{taipei}{Academia Sinica, Institute of Physics, Taipei 11529, Taiwan\Arefs{tw}}
\item \Idef{telaviv}{Tel Aviv University, School of Physics and Astronomy, 69978 Tel Aviv, Israel\Arefs{t}}
\item \Idef{triest_u}{University of Trieste, Dept.\ of Physics, 34127 Trieste, Italy}
\item \Idef{triest_i}{Trieste Section of INFN, 34127 Trieste, Italy}
\item \Idef{turin_u}{University of Turin, Dept.\ of Physics, 10125 Turin, Italy}
\item \Idef{turin_i}{Torino Section of INFN, 10125 Turin, Italy}
\item \Idef{tomsk}{Tomsk Polytechnic University,634050 Tomsk, Russia\Arefs{nauka}}
\item \Idef{illinois}{University of Illinois at Urbana-Champaign, Dept.\ of Physics, Urbana, IL 61801-3080, USA\Arefs{nsf}}
\item \Idef{warsaw}{National Centre for Nuclear Research, 00-681 Warsaw, Poland\Arefs{u}}
\item \Idef{warsawu}{University of Warsaw, Faculty of Physics, 02-093 Warsaw, Poland\Arefs{u}}
\item \Idef{warsawtu}{Warsaw University of Technology, Institute of Radioelectronics, 00-665 Warsaw, Poland\Arefs{u} }
\item \Idef{yamagata}{Yamagata University, Yamagata 992-8510, Japan\Arefs{q} }
\end{Authlist}
%%%%%%%%%%%%%%%%%%%%%%%%%%%%%%%%%%%%%%%%%%%%%%%%%%%%%%%%%%%%%%%%%%%%%%%%%%%%%%%%%%%%%%%%%%%%%%%%%%%%%%%%%%%%%%%%%%%%%%%
%
% Notes
%
%%%%%%%%%%%%%%%%%%%%%%%%%%%%%%%%%%%%%%%%%%%%%%%%%%%%%%%%%%%%%%%%%%%%%%%%%%%%%%%%%%%%%%%%%%%%%%%%%%%%%%%%%%%%%%%%%%%%%%%
\renewcommand\theenumi{\alph{enumi}}
\begin{Authlist}
\item [{\makebox[2mm][l]{\textsuperscript{\#}}}] Corresponding authors
\item [{\makebox[2mm][l]{\textsuperscript{*}}}] Deceased
\item \Adef{a}{Also at Instituto Superior T\'ecnico, Universidade de Lisboa, Lisbon, Portugal}
\item \Adef{b}{Also at Dept.\ of Physics, Pusan National University, Busan 609-735, Republic of Korea and at Physics Dept., Brookhaven National Laboratory, Upton, NY 11973, USA}
\item \Adef{ictp}{Also at Abdus Salam ICTP, 34151 Trieste, Italy}
\item \Adef{r}{Supported by the DFG cluster of excellence `Origin and Structure of the Universe' (www.universe-cluster.de) (Germany)}
\item \Adef{p2i}{Supported by the Laboratoire d'excellence P2IO (France)}
\item \Adef{d}{Also at Chubu University, Kasugai, Aichi 487-8501, Japan\Arefs{q}}
\item \Adef{x}{Also at Dept.\ of Physics, National Central University, 300 Jhongda Road, Jhongli 32001, Taiwan}
\item \Adef{e}{Also at KEK, 1-1 Oho, Tsukuba, Ibaraki 305-0801, Japan}
\item \Adef{g}{Also at Moscow Institute of Physics and Technology, Moscow Region, 141700, Russia}
%\item \Adef{v}{Supported by Presidential Grant NSh--999.2014.2 (Russia)}
\item \Adef{h}{Present address: RWTH Aachen University, III.\ Physikalisches Institut, 52056 Aachen, Germany}
\item \Adef{yerevan}{Also at Yerevan Physics Institute, Alikhanian Br. Street, Yerevan, Armenia, 0036}
\item \Adef{y}{Also at Dept.\ of Physics, National Kaohsiung Normal University, Kaohsiung County 824, Taiwan}
%\item \Adef{infn}{Also at \ref{LItriest_i})}
\item \Adef{turin_p}{Also at University of Eastern Piedmont, 15100 Alessandria, Italy}
\item \Adef{i}{Present address: Uppsala University, Box 516, 75120 Uppsala, Sweden}
\item \Adef{c}{    Supported by the DFG Research Training Group Programmes 1102 and 2044 (Germany)} %mail Horst Fischer 20/09/2016 11:08
%
% support institutes
%
\item \Adef{l}{    Supported by BMBF - Bundesministerium f\"ur Bildung und Forschung (Germany)}
\item \Adef{s}{    Supported by FP7, HadronPhysics3, Grant 283286 (European Union)}
\item \Adef{m}{    Supported by MEYS, Grant LG13031 (Czech Republic)}
\item \Adef{n}{    Supported by SAIL (CSR) and B.Sen fund (India)}
\item \Adef{o}{    Supported by CERN-RFBR Grant 12-02-91500}
\item \Adef{p}{\raggedright 
                   Supported by FCT - Funda\c{c}\~{a}o para a Ci\^{e}ncia e Tecnologia, COMPETE and QREN, Grants CERN/FP 116376/2010, 123600/2011 
                   and CERN/FIS-NUC/0017/2015 (Portugal)}
\item \Adef{q}{    Supported by MEXT and JSPS, Grants 18002006, 20540299, 18540281 and 26247032, the Daiko and Yamada Foundations (Japan)}
\item \Adef{tw}{   Supported by the Ministry of Science and Technology (Taiwan)}
\item \Adef{t}{    Supported by the Israel Academy of Sciences and Humanities (Israel)}
\item \Adef{nauka}{Supported by the Russian Federation  program ``Nauka'' (Contract No. 0.1764.GZB.2017) (Russia)}
\item \Adef{nsf}{  Supported by the National Science Foundation, Grant no. PHY-1506416 (USA)}
\item \Adef{u}{    Supported by NCN, Grant 2015/18/M/ST2/00550 (Poland)}
\end{Authlist}

}
\newpage
%\linenumbers
\setcounter{page}{1}
\parindent=0em

%%%%%%%%%%%%%%%%%%
\section{\large{Introduction}}
\label{intro}
\vspace{0.3cm}

A complete understanding of the three-dimensional parton structure of a fast moving nucleon requires the knowledge of the intrinsic motion of quarks in the plane transverse to the direction of motion, both in momentum and coordinate space. While the spatial distributions of quarks in the transverse plane are described by generalised parton distributions (GPDs), the momentum distributions of quarks in the transverse plane are described by transverse-momentum-dependent (TMD) parton distribution functions (PDFs). A precise knowledge of TMD-PDFs is found~\cite{Anselmino:2005nn} to be crucial in the explanation of many single-spin effects observed in hard scattering reactions~\cite{Adams:1991rw,Adams:2003fx,Airapetian:2001eg}, in addition to the important role they play in spin-independent processes. In a similar way, transverse-momentum-dependent fragmentation functions (TMD-FFs) are crucial for the description of hard scattering reactions involving hadron production. Both PDFs and FFs are non-perturbative quantities that are assumed to be process-independent. The simplest examples are the spin-averaged TMD-PDF $f_{1}^{q}(x,~k_{\rm{T}}$) and the spin-averaged TMD-FF $D_{q}^{h} (z,~p_{\rm{h}\perp})$, where $x$ is the Bjorken scaling variable, $k_{\rm{T}}$ is the quark intrinsic transverse momentum, $z$ is the fractional energy of the final-state hadron, and $p_{\rm{h}\perp}$ is the transverse momentum of the final-state hadron relative to the direction of the fragmenting quark. After integration over $k_{\rm{T}}$ and $p_{\rm{h}\perp}$, the TMD-PDFs and TMD-FFs reduce to the standard spin-averaged collinear PDFs and FFs, where collinear means along the direction of the virtual photon.
While the knowledge on the collinear PDFs and FFs is quite advanced, very little is presently known about the dependence of TMD-PDFs and TMD-FFs on $k_{\rm{T}}$ and $p_{\rm{h}\perp}$, as only sparse experimental data are available to date.

One of the most powerful tools to assess TMD-PDFs and TMD-FFs is the semi-inclusive measurement of deep inelastic scattering (SIDIS), $\ell N \rightarrow \ell' hX$, where one hadron is detected in coincidence with the scattered lepton in the final state. According to the QCD factorisation theorem \cite{Collins:1996fb,Ji:2004xq} the deep inelastic scattering (DIS) process is considered to proceed via two independent sub-processes, i.e. the elementary QED process $\ell  q \rightarrow \ell  q$ is followed by the hadronisation of the struck quark. 
The outgoing hadrons provide information about the original transverse motion of the quark in the nucleon via their transverse momentum vector $\textbf{\textit{P}}_{\rm{hT}}$. The latter is defined with respect to the virtual photon direction.
The SIDIS cross section can be written as a convolution of a `hard' scattering cross section, which is calculable in perturbative QCD (pQCD), with the non-perturbative TMD-PDFs and TMD-FFs. It depends on five kinematic variables. Two variables describe inclusive DIS, i.e. the negative square of the four-momentum transfer $Q^{2}=-q^{2}$ and the Bjorken scaling variable $x= - q^2/(2 P \cdot q)$, where $q$ and $P$ denote the four-momenta of the virtual photon and the nucleon, respectively. Three more variables describe the final-state hadrons, i.e. the fraction of the virtual photon energy that is carried by a hadron, $z = (P\cdot P_{\rm{h}})/(P\cdot q)$, the magnitude $P_{\rm{hT}}$ of the transverse momentum of a hadron and its azimuthal angle $\phi$ in the system of virtual photon and nucleon. Here, $P_{\rm{h}}$ denotes the four-momentum of the hadron. In the present analysis, the dependence on $\phi$ is disregarded. When integrating it over, the differential cross section for spin-independent  SIDIS reads as follows in twist two ``TMD factorisation scheme"~\cite{Bacchetta:2006tn, Anselmino:2013lza}:

\begin{equation}
\frac{\dd^4 \sigma^{\ell p\rightarrow \ell' hX}}{\dd x \dd Q^{2} \dd z\dd P^{2}_{\rm{hT}}} = \frac{2\pi^{2}\alpha^2}{(x s)^{2}} \frac{[1 + (1-y)^2]}{y^2} F_{UU}(x,~Q^2,~z,~P_{\rm{hT}}^{2}),
\end{equation}

with 
\begin{equation}
F_{UU} (x,~Q^2,~z,~P_{\rm{hT}}^{2}) = \sum_{q} e_{q}^{2} \int \dd^2 \textbf{\textit{k}}_{\rm{T}} \dd^2 \textbf{\textit{p}}_{\rm{h}\perp} \delta^{(2)}(\textbf{\textit{P}}_{\rm{hT}} - z \textbf{\textit{k}}_{\rm{T}} - \textbf{\textit{p}}_{\rm{h}\perp}) f_{1}^{q}(x,~Q^{2},~k_{\rm{T}}) D_{q}^{h}(z,~Q^{2},~p_{\rm{h}\perp}).
\label{fact_theor}
\end{equation}

Here, $y$ is the lepton energy fraction that is carried by the virtual photon and $s$ is the centre-of-mass energy, which are related to $x$ and $Q^2$ through $Q^2 = xys$.
The hadron transverse momentum is related to $\textbf{\textit{k}}_{\rm{T}}$ and $\textbf{\textit{p}}_{\rm{h}\perp}$  by $\textbf{\textit{P}}_{\rm{hT}} = z\textbf{\textit{k}}_{\rm{T}} + \textbf{\textit{p}}_{\rm{h}\perp}$~\cite{Bacchetta:2006tn}.
An important consequence of the factorisation theorem is that the fragmentation function is independent of $x$, and the parton distribution function is independent of $z$, while both depend on $Q^2$.

In addition to azimuthal asymmetries in spin-independent SIDIS~\cite{Adolph:2014pwc}, the most relevant experimental observable to investigate spin-averaged TMD-PDFs and TMD-FFs is the differential hadron multiplicity as a function of $P_{\rm{hT}}^{2}$, which is defined in Eq.~\ref{MulDef} below.
`Soft' non-perturbative processes are expected to generate relatively small values of $P_{\rm{hT}}$ with an approximately Gaussian distribution in $P_{\rm{hT}}$~\cite{Anselmino:2006rv}. Hard QCD processes are expected to generate large non-Gaussian tails for $P_{\rm{hT}} > $1~(GeV/$c$). They are expected to play an important role in the interpretation of the results reported here, which reach values of $P_{\rm{hT}}^{2}$ up to 3~(GeV/$c$)$^2$.

Transverse-momentum-dependent distributions of charged hadrons in DIS were first measured by the EMC collaboration~\cite{Ashman:1991cj} at CERN, followed by measurements by ZEUS~\cite{Derrick:1995xg} and H1~\cite{Adloff:1996dy,Aaron:2008ad} at HERA. These measurements only provided data in a limited dimensional space. Only new-generation experiments provided higher statistics, thereby opening the way to analyse and present the results in several dimensions simultaneously. Recent results were obtained by several fixed-target experiments using various targets and complementary energy regimes, i.e. HERMES~\cite{Airapetian:2012ki} at DESY and COMPASS~\cite{Adolph:2013stb} at CERN.

The present paper reports on a new COMPASS measurement of transverse-momentum-dependent multiplicities of charged hadrons and extends the results of our earlier publication on transverse-momentum-dependent  distributions of charged hadrons~\cite{Adolph:2013stb}. The present measurement enlarges the kinematic coverage in $x$ up to 0.4 instead of 0.12, in $Q^2$ up to 81~(GeV/$c$)$^2$ instead of 10~(GeV/$c$)$^2$ and in  $P_{\rm{hT}}^{2}$ up to 3~(GeV/$c$)$^2$ instead of about 1~(GeV/$c$)$^2$ with significantly reduced systematic uncertainties on the normalisation of the $P_{\rm{hT}}^{2}$-integrated multiplicities~\cite{Adolph:2013stb}. 
The data reported here represent the most precise results on differential charged hadron multiplicities available at this energy scale. 
This measurement is unique as its high statistics allows us to analyse the $P_{\rm{hT}}^{2}$-dependence of charged-hadron multiplicities in four variables simultaneously.
 
The paper is organised as follows. Section \ref{apparatus} briefly describes the experimental apparatus. Details about the data analysis are given in Sec.~\ref{dataanalysis}. The measured charged-hadron multiplicities are presented and compared to previous measurements in Sec.~\ref{results}. In Sec.~\ref{fitsection} fits to the results are presented and discussed. The results are summarised in Sec.~\ref{sumary}.

\section{\large{Experimental setup}}
\vspace{0.3cm}
\label{apparatus}

The set-up of the COMPASS experiment is shortly described in this section. A more detailed description can be found in Ref.~\cite{Abbon:2007pq}. It is a fixed-target experiment, which uses the CERN Super Proton Synchrotron M2 beamline that is able to deliver high-energy hadron and muon beams. 
The data were collected in 2006 using a naturally polarised $\mu^{+}$ beam of 160~GeV/$c$ with a momentum spread of 5\%. The intensity was $4\times10^7$~s$^{-1}$ with a spill length of 4.8~s and a cycle time of 16.8~s. The momentum of each incoming muon was measured before the COMPASS experiment with a precision of 0.3\%. The trajectory of each incoming muon was measured before the target in a set of silicon and scintillating fibre detectors. The muons were impinging on a longitudinally polarised solid-state target located inside a superconducting magnet. The target consisted of three cells that were located along the beam one after the other. It was filled with $^6$LiD beads immersed in a liquid $^3$He/$^4$He mixture. The admixtures of H, $^3$He and $^7$Li in the target lead to an effective excess of neutrons of about $0.2\%$. To first approximation, it can be regarded as an isoscalar deuteron target and will be referred to as such in the following. The polarisation of the middle cell (60~cm length) was opposite to that of the two outer cells (30~cm long each), and the polarisation was reversed once per day. In order to obtain spin-independent results, the target polarisation was averaged over by combining the data from all three target cells. Since the data taking in the two target polarisation states was well balanced and remaining polarisation-dependent effects are very small, this procedure ensures that for the data analysis the target can be considered as unpolarised. 

The COMPASS two-stage spectrometer was designed to reconstruct scattered muons and produced hadrons in a wide range of momentum and polar angle, where the latter reaches up to 180 mrad. Particle tracking is performed by a variety of tracking detectors that are located before and after the two spectrometer magnets. The direction of the reconstructed tracks at the interaction point is determined with a precision of 0.2~mrad. The momentum resolution is 1.2\% in the first spectrometer stage and 0.5\% in the second one. The trigger is made by hodoscope systems supplemented by hadron calorimeters. Muons are identified downstream of hadron absorbers.

%~~~~~~~~~~~~~~~~~~~~~~~~~~~~~~~~~~~~~~~~~~~~~~%
%                                                                                                 %
%                    			New section                                       %
%                                                                                                 %
%~~~~~~~~~~~~~~~~~~~~~~~~~~~~~~~~~~~~~~~~~~~~~~%
\section{\large{Multiplicity and data analysis}}
%\vspace{0.3cm}
\label{dataanalysis}

%~~~~~~~~~~~~~~~~~~~~~~~~~~~~~~%
%                                                               %
%                  subsection n.1                      %
%                                                               %
%~~~~~~~~~~~~~~~~~~~~~~~~~~~~~~%

\subsection{\large{Multiplicity extraction}}
\label{ss0}
\vspace{0.3cm}

The differential multiplicity $M^{\rm{h}}$ for charged hadrons, where h denotes a long-lived charged hadron ($\pi^{+}$, $\pi^{-}$, K$^{+}$, K$^{-}$, p or $\bar{\rm{p}}$), is defined as the ratio between the differential semi-inclusive cross section $\dd^4\sigma^{\rm{h}}$ and the differential inclusive cross section $\dd^2\sigma^{\rm{DIS}}\colon$

\begin{equation}
 \frac{\dd^2M^{\rm{h}}(x,Q^2,z,P_{\rm{hT}}^{2})}{\dd z\dd P_{\rm{hT}}^{2}}= \left( \frac{\dd^4\sigma^{\rm{h}}}{\dd x\dd Q^2\dd z\dd P_{\rm{hT}}^{2}}\right) \Bigg/ \left( \frac{\dd^2 \sigma^{\rm{DIS}}}{\dd x\dd Q^2}\right).
\label{MulDef}
\end{equation}

 Hadron multiplicities are measured in the four-dimensional ($x$,~$Q^2$,~$z$,~$P_{\rm{hT}}^{2}$) space. The bin limits in the four variables are presented in Table~\ref{binning}. 

The data used in the present analysis were collected during six weeks in 2006. The data analysis comprises event and hadron selection, the correction for radiative effects, the determination of and the correction for the kinematic and geometric acceptance of the experimental set-up as well as for detector inefficiencies, detector resolutions and bin migration, and the correction for diffractive vector-meson production. Differential hadron multiplicities are evaluated as the ratio of hadron yields $\dd^4 N^{\rm{h}}$ in every interval of ($x,~Q^2,~z,~P_{\rm{hT}}^2$) and the number of DIS events $\dd^2 N^{\rm{DIS}}$ in every interval of ($x,~Q^2$) corrected as described above$\colon$

\begin{equation}
\frac{\dd^2M^{\rm{h}}(z,~P_{\rm{hT}}^{2})}{\dd z\dd P_{\rm{hT}}^{2}} = \frac{1}{\dd^2N^{\rm{DIS}}}\frac{\dd^4 N^{\rm{h}}(z,~P_{\rm{hT}}^{2})}{\dd z\dd P_{\rm{hT}}^{2}}\left( 1-\frac{\eta^{\rm{h}}}{\eta^{\rm{DIS}}} \right) \frac{1}{a^{\rm{h}}(z,~P_{\rm{hT}}^{2})}\frac{C^{\rm{h}}(z,~P_{\rm{hT}}^{2})}{C^{\rm{DIS}}}.
\label{mul}
\end{equation}

Here, $\eta^{\rm{DIS}}$ and $\eta^{\rm{h}}$ denote the correction factors accounting for radiative effects in the inclusive and in the semi-inclusive case, respectively, $a^{h}$ accounts for acceptance effects, and $C^{\rm{DIS}(h)}$ denotes the correction factor accounting for the diffractive vector-meson contribution in the case of an inclusive (semi-inclusive) measurement. The $(x,~Q^2)$ dependence is omitted for simplicity as it enters all terms. 
All corrections are evaluated in the four-dimensional $(x,~Q^2,~z,~P_{\rm{hT}}^{2})$ bins except $C^{\rm{DIS}}$, $\eta^{\rm{DIS}}$ and $\eta^{\rm{h}}$, which are evaluated only in bins of $x$ and $Q^2$. Further kinematic dependences of $\eta^{\rm{h}}$ upon $z$ and $P_{\rm{hT}}^{2}$ will be discussed in Sec.~\ref{ss1}.

\begin{table}[htp]
\caption{Bin limits for the four-dimensional binning in $x$,~$Q^2$,~$z$ and $P_{\rm{hT}}^{2}$.}
\begin{center}
\begin{tabular}{p{2.5cm}p{1cm}p{1cm}p{1cm}p{1cm}p{1cm}p{1cm}p{1cm}p{1cm}p{1cm}} \hline 
 & \multicolumn{9}{l}{bin limits}\\ \hline
$x$ 					& 0.003 & 0.008 & 0.013 & 0.02 & 0.032 & 0.055 & 0.1 & 0.21 & 0.4\\
$Q^2$~(GeV/$c$)$^2$ 	& 1.0 & 1.7 & 3.0 & 7.0 & 16 & 81 & & & \\
$z$ 					& 0.2  & 0.3 & 0.4 & 0.6 & 0.8 & & & & \\
$P_{\rm{hT}}^{2}$~(GeV/$c$)$^2$	& 0.02 & 0.04 & 0.06 & 0.08 & 0.10 & 0.12 & 0.14 & 0.17 & 0.196 \\
						& 0.23 & 0.27 & 0.30 & 0.35 & 0.40 & 0.46 & 0.52 & 0.60 & 0.68\\
						& 0.76 & 0.87 & 1.00 & 1.12 & 1.24 & 1.38 & 1.52 & 1.68 & 1.85\\
						& 2.05 & 2.35 & 2.65 & 3.00 & & & & & \\
 \hline
\end{tabular}
\end{center}
\label{binning}
\end{table}

%~~~~~~~~~~~~~~~~~~~~~~~~~~~~~~%
%                                                               %
%                  subsection n.2                      %
%                                                               %
%~~~~~~~~~~~~~~~~~~~~~~~~~~~~~~%
\subsection{\large{Event and hadron selection}}
\label{ss1}
\vspace{0.3cm}

\begin{figure}[htp]
\centering
\includegraphics[height=6.5cm,scale=0.4]{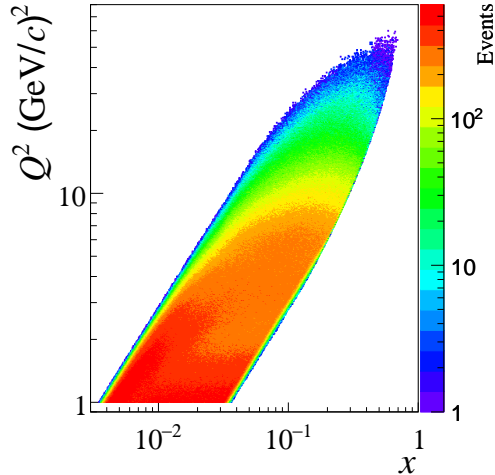}
\caption{Distribution and kinematic range of the selected DIS sample in the ($x$,~$Q^2$) plane.}
\label{xq2dis}
\end{figure}

The present analysis uses events taken with 'inclusive triggers', i.e. the trigger decision is based on scattered muons only. The selected events are required to have a reconstructed interaction vertex associated with an incident and a scattered muon track. This vertex has to lie inside a fiducial target volume. The incident muon energy is constrained to the range from 140 GeV to 180 GeV. In addition to the kinematic constraints given by the spectrometer acceptance, the selected events are required to have $Q^{2}>$1~(GeV/c)$^2$ and $W>5$~GeV/$c^2$. These requirements select the DIS regime and exclude the nucleon resonance region. The relative virtual-photon energy is constrained to the range $0.1 < y < 0.9$ to exclude kinematic regions where the momentum resolution degrades and radiative effects are most pronounced~\cite{Makke:2011fia}. In the range $0.003 < x < 0.4$, the total number of inclusive DIS events is $13\times10^{6}$, which corresponds to an integrated luminosity of 0.54~fb$^{-1}$. The ($x$,~$Q^2$) distribution of this selected `DIS sample' is shown in Fig.~\ref{xq2dis}, where a strong correlation between $x$ and $Q^{2}$ is observed as expected in fixed-target experiments.

For a selected DIS event, all reconstructed tracks associated with the primary interaction vertex are considered. Hadron tracks must be detected in detectors located before and after the magnet in the first stage of the spectrometer. The fraction of the virtual-photon energy transferred to a final-state hadron is constrained to $0.2 < z < 0.8$. The lower limit excludes the target fragmentation region, while the upper one removes muons wrongly identified as hadrons and excludes the region with larger contributions from diffractive $\rho^{0}$ production. This selection yields the `hadron sample' with a total of $4.3\times 10^{6}$ and $3.4 \times 10^{6}$ positively and negatively charged hadrons, respectively.

The corrections for QED higher-order effects are applied on an event-by-event basis taking into account the target composition. They are computed as a function of $x$ and $y$ according to the scheme described in Ref. \cite{Akhundov:1994my}. For the hadron yields, the correction is calculated by excluding the elastic and quasi-elastic tails. The correction factors $\eta^{\rm{h}}$ and $\eta^{\rm{DIS}}$ are evaluated in bins of $x$ and $Q^2$. They are found to be smaller than 12 \% for $x < 0.01$ and are smaller than 5\% elsewhere.
An attempt to evaluate the smearing due to radiative effects as a function of $z$ and $P_{\rm{hT}}^{2}$ was performed using a Monte Carlo (MC) simulation, where radiative effects were simulated using the RADGEN generator \cite{Akushevich:1998ft}. A possible impact on the ($z$,~$P_{\rm{hT}}^{2}$) dependence of the results due to radiative effects is accounted for in the systematic uncertainties of the $P_{\rm{hT}}^{2}$-dependence of the multiplicities.

%~~~~~~~~~~~~~~~~~~~~~~~~~~~~~~%
%                                                               %
%                  subsection n.3                      %
%                                                               %
%~~~~~~~~~~~~~~~~~~~~~~~~~~~~~~%

\subsection{\large{Acceptance correction}}
\label{accep}
\vspace{0.3cm}

The hadron multiplicities must be corrected for geometric and kinematic acceptances of the experimental set-up as well as for detector inefficiencies and resolutions, and for bin migration.  The correction for a possible misidentification of electrons as hadrons is included in the acceptance correction. The full correction factor is evaluated using a MC simulation of the muon-deuteron deep inelastic scattering processes. Events are generated using the LEPTO generator~\cite{Ingelman:1996mq}, where the parton hadronisation mechanism is simulated using the JETSET package~\cite{Sjostrand:1995iq} with the tuning from Ref.~\cite{Adolph:2012vj}. Secondary hadron interactions are simulated using the FLUKA package~\cite{Ferrari:2005zk}. The experimental set-up is simulated using the GEANT3 toolkits~\cite{Brun:1985ps} and the MC data are reconstructed using the same software that was used for the experimental data~\cite{Abbon:2007pq}. The kinematic distributions of the experimental data are fairly well reproduced by the MC simulation.

In order to minimise a possible dependence on the physics generator used in the simulation and to exclude kinematic regions with large acceptance corrections, a four-dimensional evaluation of the acceptance correction factor $a^{\rm{h}}$ is performed in narrow kinematic bins. In each ($x_{\rm{r}}$,~$Q^2_{\rm{r}}$) kinematic bin, where r denotes the reconstructed values of the variables, the acceptance correction is calculated as the ratio of reconstructed ($\dd^2 N^{\rm{h}}_{\rm{r}}$) and generated ($\dd^2N^{\rm{h}}_{\rm{g}}$) hadron yields, where both are evaluated using the simulated DIS sample after reconstruction$\colon$

\begin{equation}
 a^{\rm{h}}(z,~P_{\rm{hT}}^{2}) = \left( \frac{\dd^2 N^{\rm{h}}_{\rm{r}}}{\dd z\dd P_{\rm{hT}}^{2}}\right) \Bigg/ \left( \frac{\dd^2N^{\rm{h}}_{\rm{g}}}{\dd z\dd P_{\rm{hT}}^{2}}\right)
\end{equation}

\begin{figure}[htp]
\centering
\subfigure[]{\label{rho}\includegraphics[height=6.2cm,width=.4\textwidth]{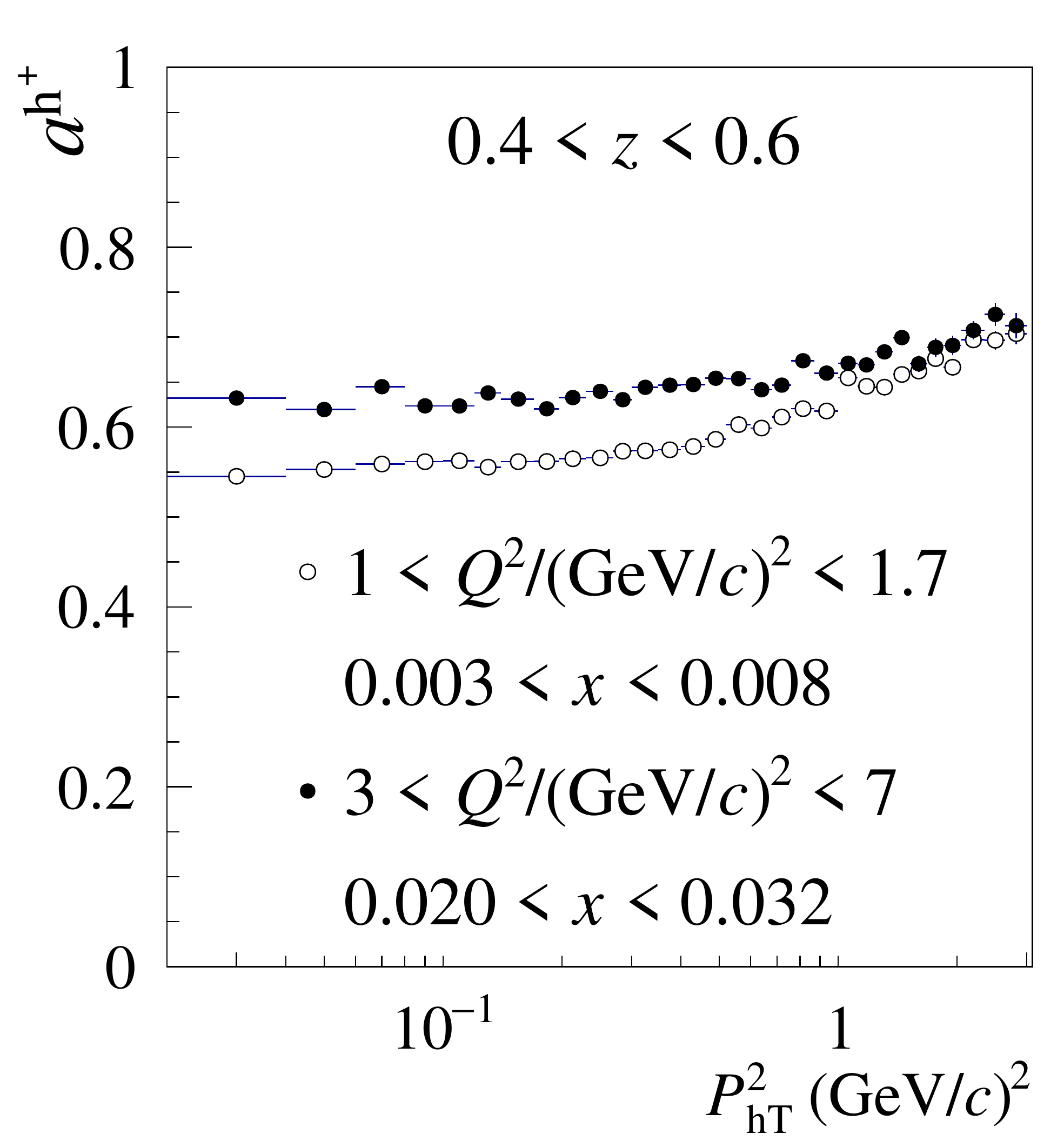}}
\subfigure[]{\label{phi}\includegraphics[height=6.2cm,width=.4\textwidth]{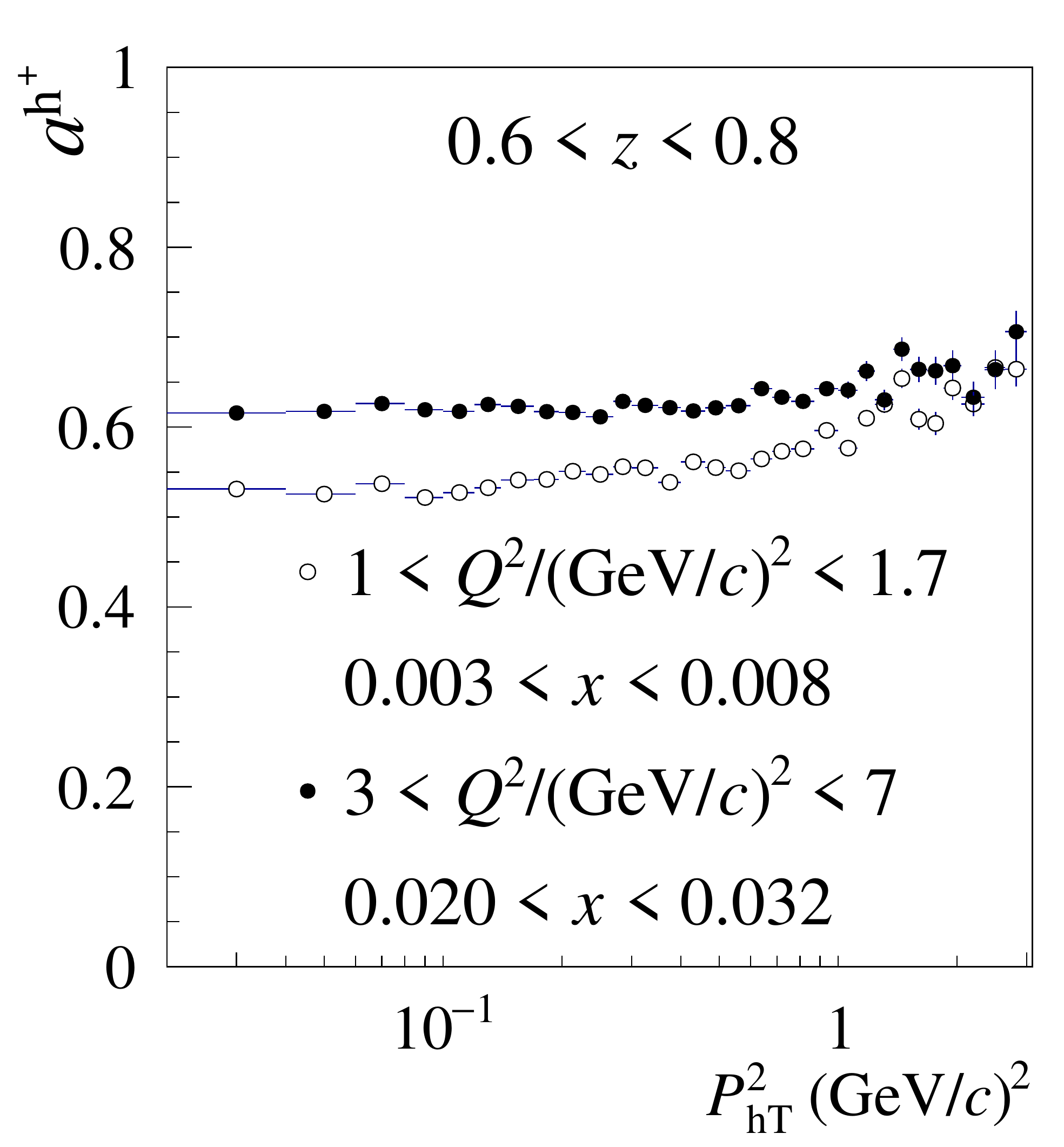}}
\caption{(a) Acceptance for positively charged hadrons as a function of $P_{\rm{hT}}^2$ for $0.4 < z < 0.6$ in two typical ($x$,~$Q^2$) bins. (b) Same as (a) for $0.6 < z < 0.8$.}
\label{acchp}
\end{figure}

An advantage of this definition is that the correction for muon acceptance cancels as it enters both numerator and denominator.
The generated values of kinematic variables are used for the generated particles and the reconstructed values of kinematic variables are used for the reconstructed particles. All reconstructed MC events and particles are subject to the same kinematic and geometric selection criteria as the data, while the generated ones are subject to kinematic requirements only. The acceptance correction factor exhibits an almost flat behaviour as a function of $z$ and $P_{\rm{hT}}^{2}$ in most ($x$,~$Q^2$) bins, except at high $x$ for $P_{\rm{hT}}^{2} > 1$~(GeV/$c$)$^2$, where it remains larger than 0.4. Elsewhere, its average value is above and close to 0.6 for $P_{\rm{hT}}^{2} > 0.5$~(GeV/$c$)$^2$ and is less than or equal to 0.6 for $P_{\rm{hT}}^{2}<0.5$~(GeV/$c$)$^2$. 
As an example, Fig.~\ref{acchp} shows the acceptance as a function of $P_{\rm{hT}}^{2}$ for positively charged hadrons. The two panels show the two $z$ bins between 0.4 and 0.8, with two bins in ($x$,~$Q^2$) in each case. The acceptance correction factors for positively and negatively charged hadrons are found to be very similar, with differences on the level of 0.02-0.04.

%~~~~~~~~~~~~~~~~~~~~~~~~~~~~~~%
%                                                               %
%                  subsection n.4                      %
%                                                               %
%~~~~~~~~~~~~~~~~~~~~~~~~~~~~~~%

\subsection{\large{Diffractive vector meson contribution}}
\vspace{0.3cm}

The final-state hadron(s) selected as described above may also originate from diffractive production of vector mesons ($\rho^0$, $\phi$, $\omega$) that decay into lighter hadrons ($\pi$,~K,~p)~\cite{Airapetian:2001eg,Adolph:2016bga,Adolph:2016bwc}. This process, which can be described by the fluctuation of the virtual photon into a vector meson that subsequently interacts diffractively with the nucleon through multiple gluon exchange, is different from the interaction of the virtual photon with a single quark in the DIS process. The fraction of selected final-state hadrons originating from diffractive vector-meson decays and their contribution to the SIDIS yields are estimated in each kinematic bin using two Monte Carlo simulations. The first one uses the LEPTO generator to simulate SIDIS events, and the other one uses the HEPGEN generator~\cite{Sandacz:2012at} to simulate diffractively produced $\rho^0$ and $\phi$ events. Further channels, which are characterised by smaller cross sections, are not taken into account. Events with diffractive dissociation of the target nucleon represent about 25\% of those with the nucleon staying intact and are also simulated. The simulation of these events includes nuclear effects, i.e. coherent production and nuclear absorption as described in Ref.~\cite{Sandacz:2012at}. 
The contribution of pions originating from $\rho^0$ decay to the hadron sample increases with $z$, and reaches up to $40-50$\% for $z$ close to 1. For kaons, the contribution from $\phi$ decay is concentrated in the $z$ range $0.4-0.6$, where it reaches up to 15\%.
The correction factors are separately evaluated for the DIS sample and the hadron sample$\colon$

\begin{equation}
C^{\rm{DIS}}(x,~Q^2) = 1-f_{\rm{DIS}}^{\rm{VM}}(x,Q^2),
\label{}
\end{equation}

\begin{equation}
C^{\rm{h}}(x,~Q^2,~z,~P_{\rm{hT}}^{2}) = \left[ F_{\pi}(1 - f_{\pi}^{\rho^0}) + F_{K}(1 - f_{\rm{K}}^{\phi}) + F_{\rm{p}} \right].
\label{}
\end{equation}

Here, $f_{\rm{DIS}}^{\rm{VM}}$ denotes the fraction of diffractively produced vector-mesons present in the DIS sample, while $f_{\pi}^{\rho^{0}}$ and $f_{\rm{K}}^{\phi}$ denote the fraction of $\rho^{0}$ and $\phi$ decay products in the hadron sample, respectively. The fraction of pions, kaons, and protons in the latter sample, which is denoted by $F_{\pi,~\rm{K},~\rm{p}}$, amounts to about 75\%, 20\% and 5\%. The fractions $F_{i}$ ($i=\pi,~\rm{K},~\rm{p}$) and $f_{i}^{\rm{VM}}$ ($i=\pi,~\rm{K}$ and $\rm{VM}=\rho^{0}$,~$\phi$) are evaluated as functions of $x$, $Q^2$, $z$ and $P_{\rm{hT}}^{2}$. 

\begin{figure}[htp]
\centering
\includegraphics[height=7.cm,width=.5\textwidth]{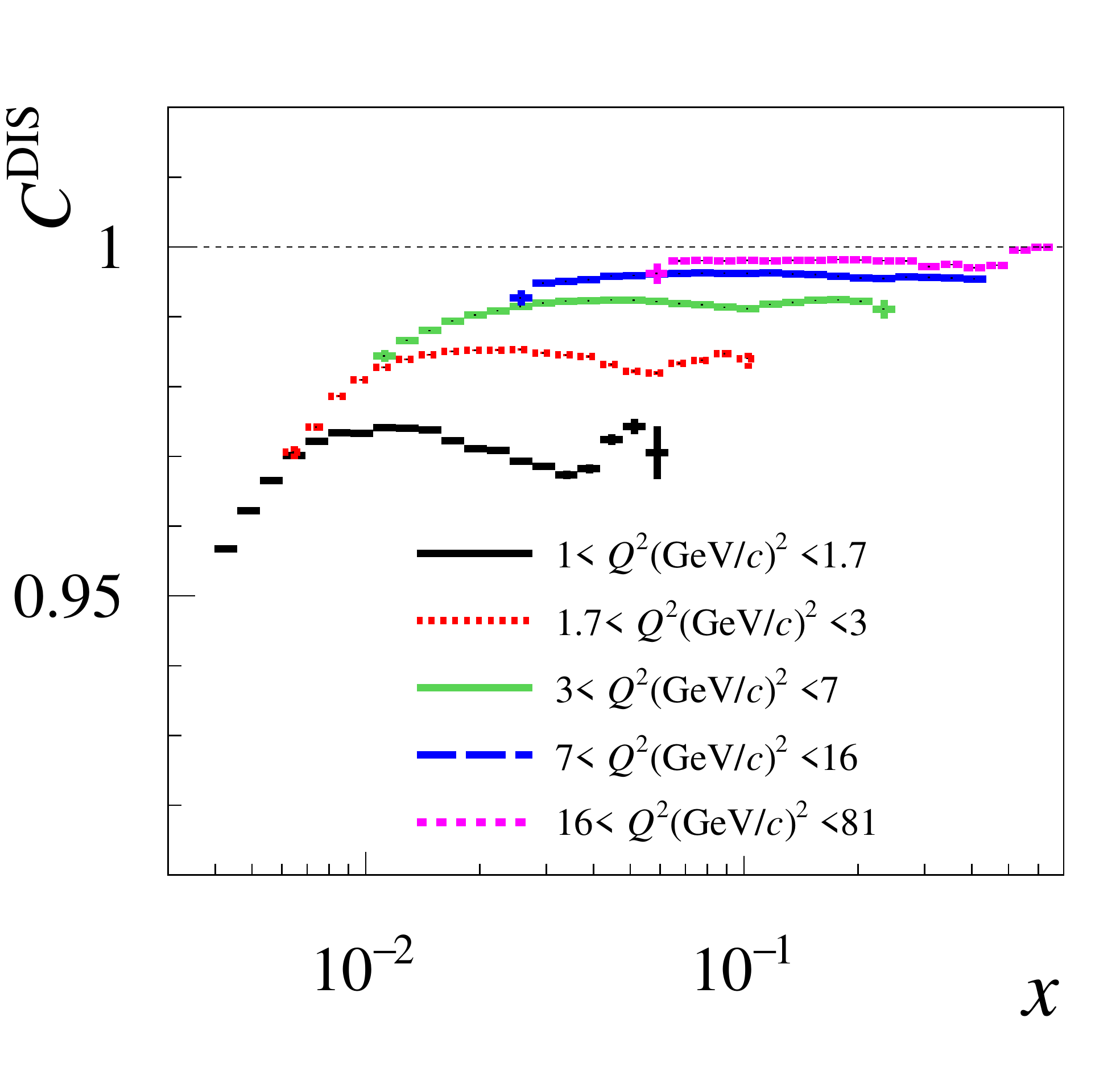}
\caption{Correction factor to the DIS yield due to diffractive $\rho^{0}$ production as a function of $x$ in the five $Q^{2}$ bins.}
\label{C_DIS}
\end{figure}

In the following, the general behaviour of some of the above discussed correction factors is illustrated.
The correction factor to account for diffractive $\rho^{0}$ production in the DIS yield is shown in Fig.~\ref{C_DIS}  as a function of $x$ in the five $Q^2$ bins. It reaches a maximum value of about 4\% in the lowest $Q^2$ bin.
The correction factor for the contribution of diffractively produced $\rho^{0}$ mesons to the pion sample, ($1-f_{\pi}^{\rho^0}$), is shown in Fig.~\ref{rho} as a function of $P_{\rm{hT}}^{2}$ in the four $z$ bins for the lowest $Q^2$ bin, where it has the largest value. It reaches a maximum value of about 25\% for $P_{\rm{hT}}^{2}\sim0.12$~(GeV/$c$)$^2$ in the highest $z$ bins, i.e. $0.6 < z < 0.8$, and decreases to few percent at small $z$. The correction factor for the contribution of diffractively produced $\phi$ mesons to the kaon sample, ($1-f_{\rm{K}}^{\phi}$), is shown in Fig.~\ref{phi}. In this case, the maximum correction of about 35\% is reached at very small $P_{\rm{hT}}^{2}$ in the middle $z$ bin, i.e. $0.4 < z < 0.6$.

\begin{figure}[htp]
\centering
\subfigure[]{\label{rho}\includegraphics[height=7.cm,width=.47\textwidth]{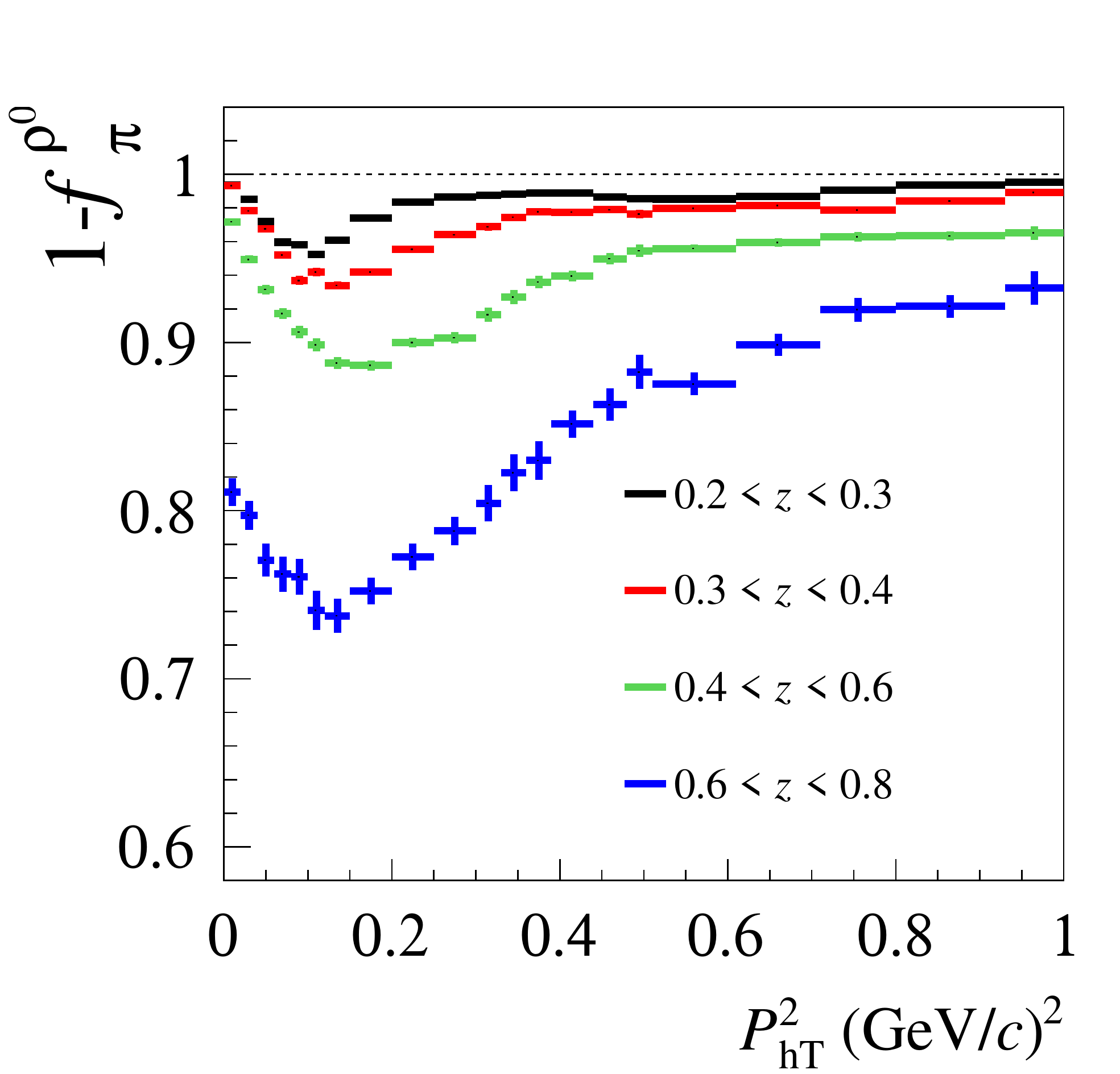}}
\subfigure[]{\label{phi}\includegraphics[height=7.cm,width=.47\textwidth]{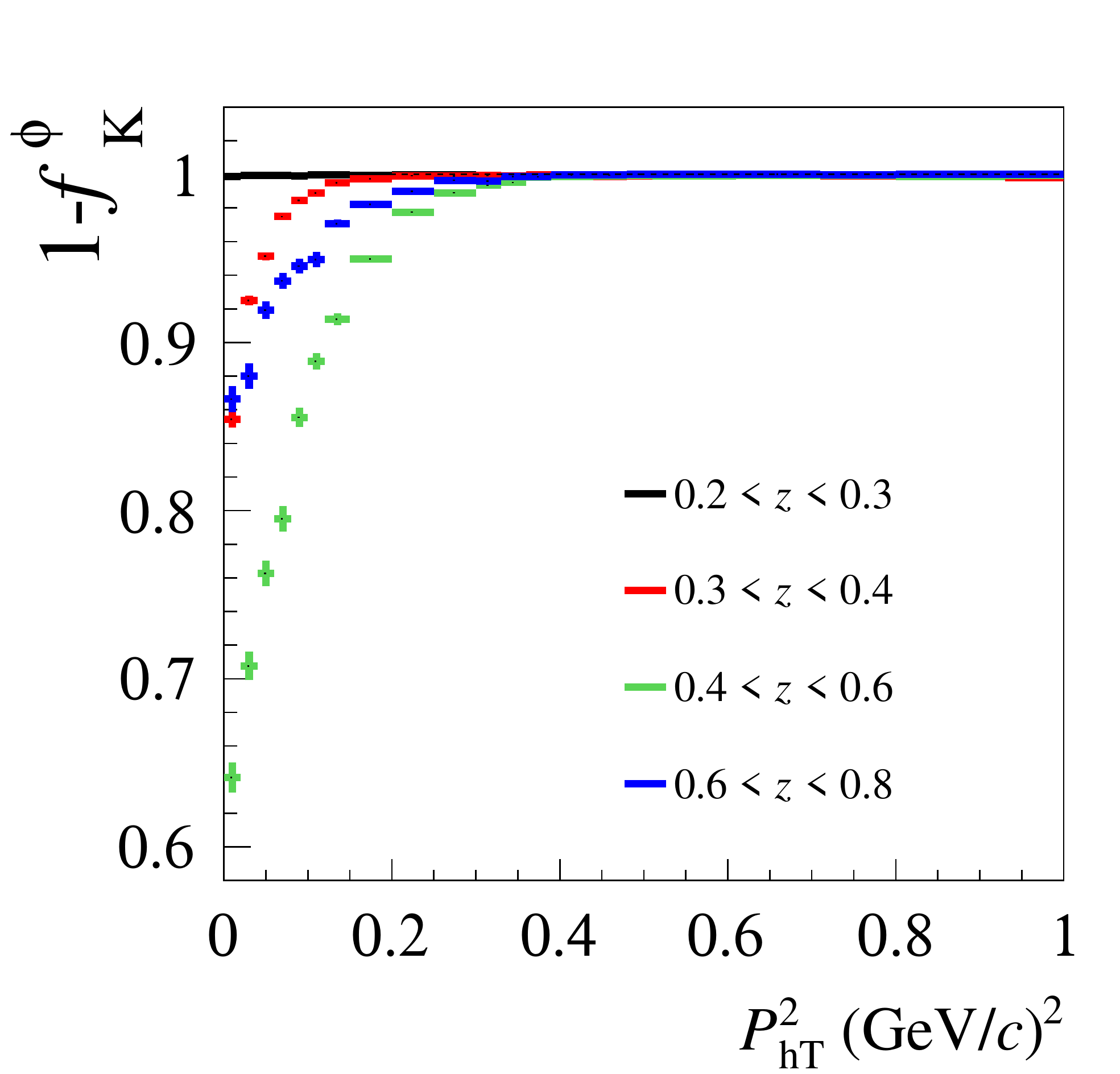}}
\caption{(a): Correction factor due to pions originating from the decay of $\rho^0$ mesons, shown as a function of $P_{\rm{hT}}^{2}$ in the four $z$ bins for $1$~(GeV/$c$)$^2 <Q^2< 1.7$~(GeV/$c$)$^2$. (b): Correction factor due to kaons originating from the decay of $\phi$ mesons.}
\label{C_SIDIS}
\end{figure}

%~~~~~~~~~~~~~~~~~~~~~~~~~~~~~~%
%                                                               %
%                  subsection n.5                      %
%                                                               %
%~~~~~~~~~~~~~~~~~~~~~~~~~~~~~~%

\subsection{\large{Systematic uncertainties}}
\vspace{0.3cm}

The dominant contributions to the systematic uncertainties originate from the uncertainties on the determination of the acceptance correction factor and those of the diffractive vector-meson contribution.
The uncertainty on the acceptance calculation is evaluated by varying in the MC simulation both the PDF set and the JETSET parameters describing the hadronisation mechanism.
The acceptance correction is estimated for each MC sample and the largest deviation with respect to the values obtained using the MC simulation described in section~\ref{accep} is quoted as a systematic uncertainty. The validity of the correction for the electron contamination is confirmed by comparing the simulated and measured electron distributions for momenta below 8 GeV/$c$, where electrons are identified using the RICH detector. 
In order to check a possible dependence on the target cell, in which the event vertex is located, the multiplicities are independently measured from the three target cells. Results from upstream and downstream target cells agree within 2-3\%, while the agreement is better than 1\% with the middle target cell. These differences are well covered in the acceptance correction uncertainty. A total uncertainty of 5\% is estimated for the multiplicities.

The cross section for exclusive production of $\rho^0$ calculated in HEPGEN is normalised to the phenomenological model of Ref.~\cite{Goloskokov:2007nt}. The theoretical uncertainty on the predicted cross section in a kinematic region close to COMPASS kinematics amounts to about 30\%. This results in an uncertainty on the diffractive vector-meson correction factor, which amounts up to $5-6\%$ mainly at small values of $x$, $Q^2$ and $P_{\rm{hT}}^{2}$, and large values of $z$.

Nuclear effects may be caused by the presence of $^3$He/$^4$He and $^6$Li in the target. The EMC Collaboration has studied in detail such nuclear effects in a similar kinematic range using carbon, copper and tin targets~\cite{Ashman:1991cj}. A $z$-dependent decrease of 5\% was observed for the multiplicities obtained using copper compared to the ones obtained using deuterium. While the effect was larger for tin, no such effect was found for carbon, so that possible nuclear effects in the present experiment are expected to be very small and are hence neglected. When comparing the results obtained from the data taken in six different weeks, no difference is observed.

All contributions to the systematic uncertainties are added in quadrature and yield a total systematic uncertainty of $5-7\%$, except at large $z$ and at large $P_{\rm{hT}}^{2}$ ($> 2.5$~(GeV/$c$)$^2$) where it reaches about 10\%. The total systematic uncertainties are shown as bands in Figs.~\ref{fullmul0} to \ref{fullmul3}. Systematic uncertainties in other figures are not shown.

%~~~~~~~~~~~~~~~~~~~~~~~~~~~~~~~~~~~~~~~~~~~~~~%
%                                                                                                 %
%                                   New section                                          %
%                                                                                                 %
%~~~~~~~~~~~~~~~~~~~~~~~~~~~~~~~~~~~~~~~~~~~~~~%
\section{\large{Measured hadron multiplicities and comparison with other experiments}}
\vspace{0.3cm}
\label{results}

%~~~~~~~~~~~~~~~~~~~~~~~~~~~~~~%
%                                                               %
%                  subsection n.1                      %
%                                                               %
%~~~~~~~~~~~~~~~~~~~~~~~~~~~~~~%

\subsection{Results}
\label{results1}
\vspace{0.3cm}

\begin{figure}[htp]
\centering
\includegraphics[height=10.cm,width=.7\textwidth]{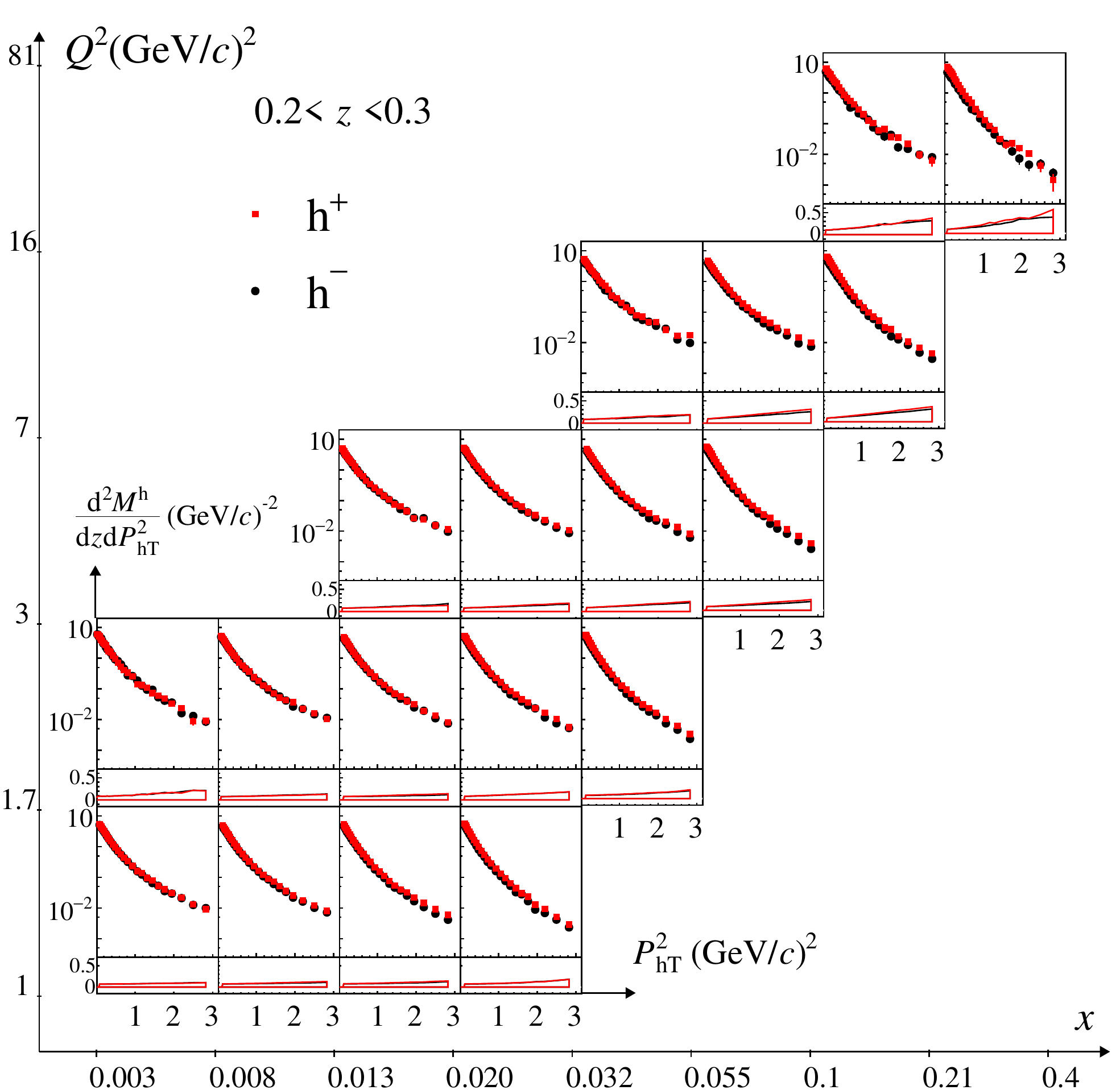}
\caption{Multiplicities of positively (full squares) and negatively (full circles) charged hadrons as a function of $P_{\rm{hT}}^{2}$ in ($x$,~$Q^{2}$) bins for $0.2<z<0.3$. Error bars on the points correspond to the statistical uncertainties. The systematic uncertainties ($\sigma_{\rm{sys}}/M^{h}$) are shown as bands at the bottom.}
\label{fullmul0}
\end{figure}

\begin{figure}[htp]
\centering
\includegraphics[height=10.cm,width=.7\textwidth]{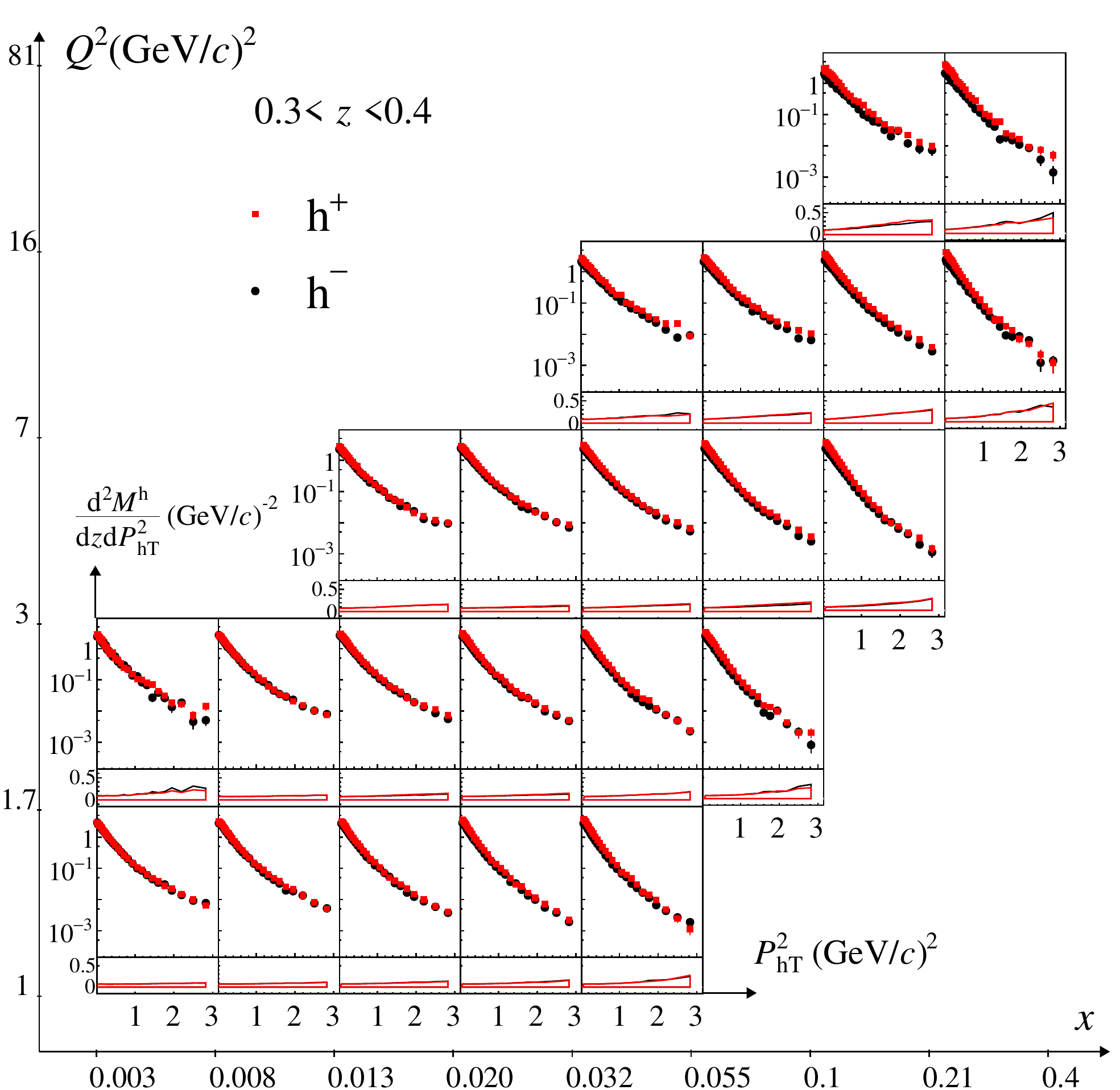}
\caption{Same as Fig.~\ref{fullmul0} for $0.3 < z < 0.4$.}
\label{fullmul1}
\end{figure}

\begin{figure}[htp]
\centering
\includegraphics[height=10.cm,width=.7\textwidth]{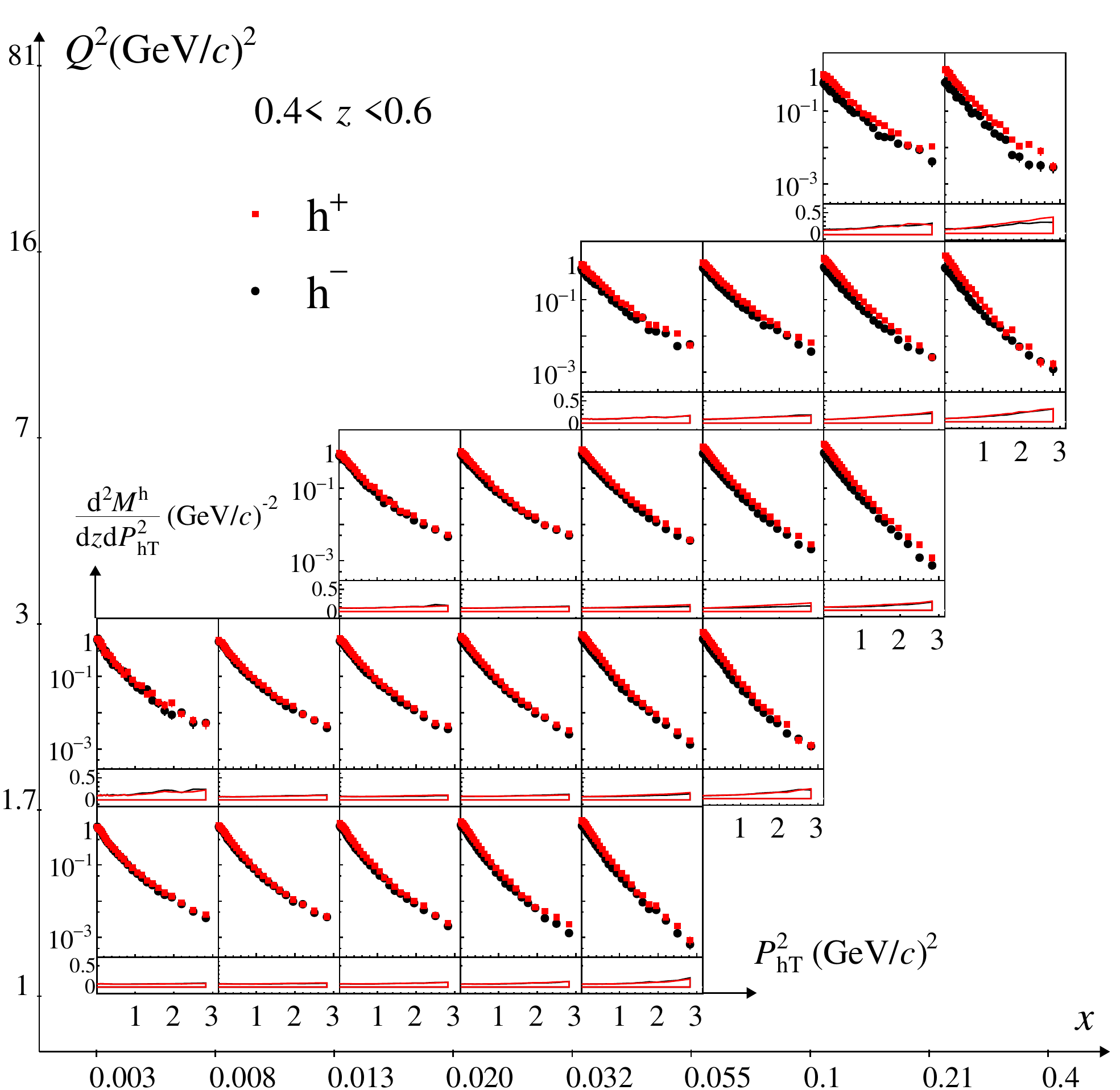}
\caption{Same as Fig.~\ref{fullmul0} for $0.4 < z < 0.6$.}
\label{fullmul2}
\end{figure}

\begin{figure}[htp]
\centering
\includegraphics[height=10.cm,width=.7\textwidth]{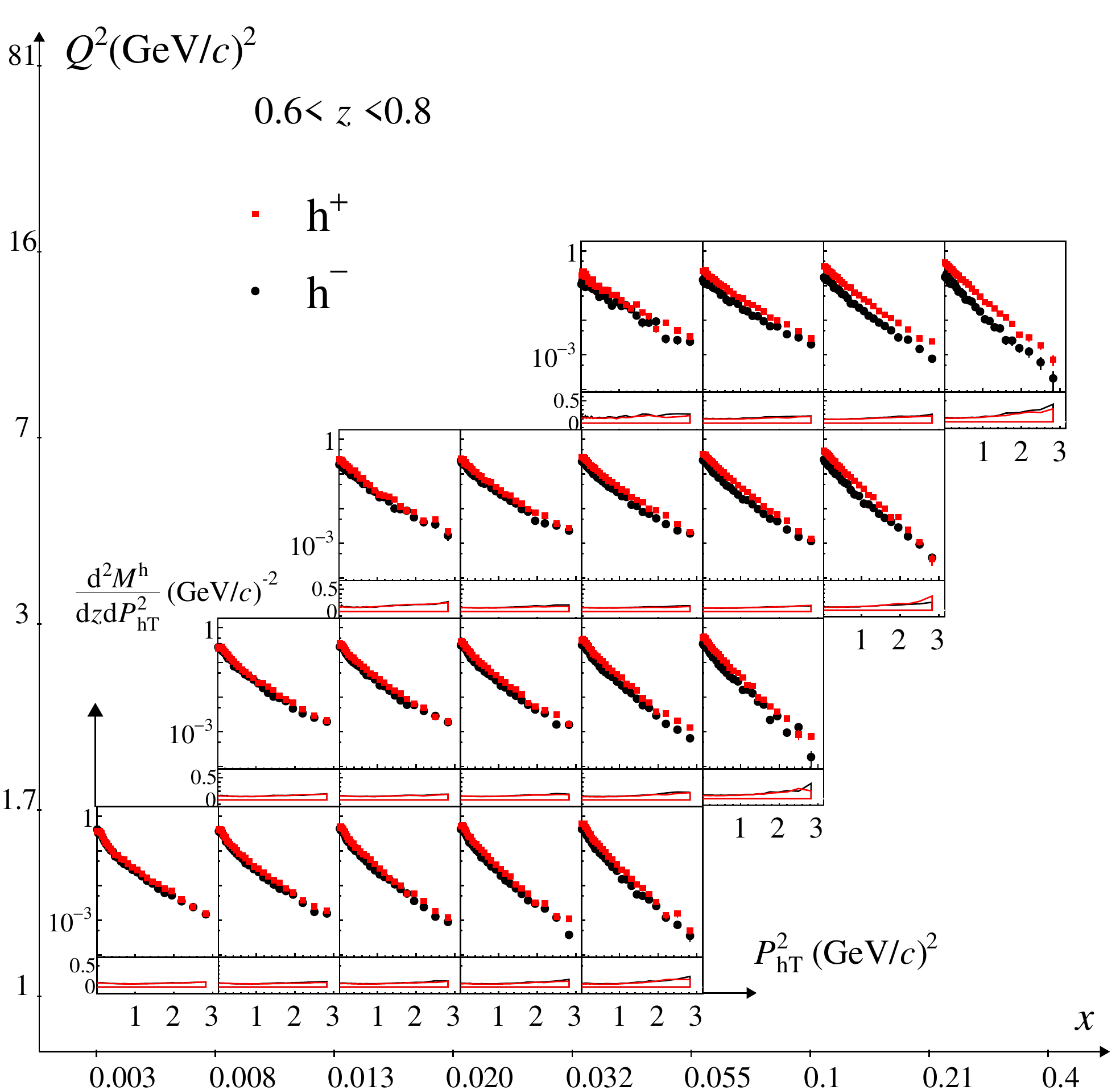}
\caption{Same as Fig.~\ref{fullmul0} for $0.6 < z < 0.8$.}
\label{fullmul3}
\end{figure}

The measured multiplicities of charged hadrons are presented in the four $z$ bins ranging from $z=0.2$ to $z=0.8$ in Figs.~\ref{fullmul0}$-$\ref{fullmul3} as a function of the hadron transverse momentum $P_{\rm{hT}}^{2}$ in bins of $x$ and $Q^{2}$. Error bars showing the statistical uncertainties on the points are too small to be visible. The systematic uncertainties are given as bands at the bottom. All multiplicities presented in the following figures are corrected for diffractive vector-meson production. The results amount to a total of 4918 experimental data points. Their numerical values are available on HepData~\cite{hepDATA} with and without correction for diffractive vector-meson production. 
It should be noted that a few ($x$, $Q^2$) kinematic bins are discarded in the lowest (Fig.~\ref{fullmul0}) and the highest (Fig.~\ref{fullmul3}) bins of $z$ because of low statistical precision as well as large acceptance correction factors (Sec.~\ref{accep}). The average values of $x$ and $Q^2$ in the various kinematic bins are evaluated using the DIS sample and are given in Table~\ref{averagevalues}.
The results obtained by integrating the multiplicities presented here over $P_{\rm{hT}}^{2}$ are in very good agreement with those of Ref.~\cite{Adolph:2016bga}, where the multiplicities of charged pions are measured as a function of $z$ in a restricted momentum range based on an independent analysis of the same data.

\begin{table}[htp]
\caption{Mean values of $x$ (left) and $Q^2$~(GeV/$c$)$^2$ (right) in all ($x$,~$Q^2$) kinematic bins defined in Tab.~\ref{binning}. For each case, rows correspond to the eight $x$ bins and columns correspond to the five $Q^2$ bins.}
\begin{center}
\begin{tabular}{|c||p{.85cm}p{.81cm}p{.6cm}p{.75cm}p{.9cm}|p{.85cm}p{.81cm}p{.6cm}p{.75cm}p{.9cm}|} \hline
$Q^{2}$ (GeV/$c$)$^2$ bins $\rightarrow$ & 
\makebox[3em]{1-1.7	} & \makebox[3em]{1.7-3}	& \makebox[3em]{3-7}	&	\makebox[3em]{7-16}	&	\makebox[3em]{16-81} &	\makebox[3em]{1-1.7} & 	\makebox[3em]{1.7-3}	&	\makebox[3em]{3-7}	&	\makebox[3em]{7-16}	&	\makebox[3em]{16-81} \\
$x$ bins $\downarrow$ &&&&&&&&&&\\ \hline

0.003 - 0.008	&	0.0062	& 	0.0074	&			&		  &		  &	1.25	&	1.82	&		&		&		\\
0.008 - 0.013 	&	0.010  	&	0.011	&			&		  &		  &	1.32	&	2.12	&		&		&		\\
0.013 - 0.02	&	0.016 	& 	0.016	&	0.017	&		  &		  &	1.30	&	2.28	&	3.51	&		&		\\
0.02   - 0.032	&	0.025 	&  	0.025	&	0.026	&		  &		  &	1.29	&	2.29	&	4.10	&		&		\\
0.032-0.055 	&	0.039 	&	0.042	&	0.043	&	0.045 &		  &	1.39	&	2.29	&	4.52	&	8.33	&		\\
0.055-0.1		&		 	&	0.068	&	0.075	&	0.077 &		  &		&	2.47	&	4.65	&	9.30	&		\\
0.1-0.21		&			& 			&	0.133	&	0.149 &	0.157 &		&		&	5.29	&	9.78	&	19.9	\\
0.21-0.4		&			&			&			&	0.254 &	0.291 &		& 		&		&	11.04 &	22.1	\\
\hline
\end{tabular}
\end{center}
\label{averagevalues}
\end{table}

\begin{figure}[!h]
\centering
\includegraphics[height=10.2cm,width=0.6\textwidth]{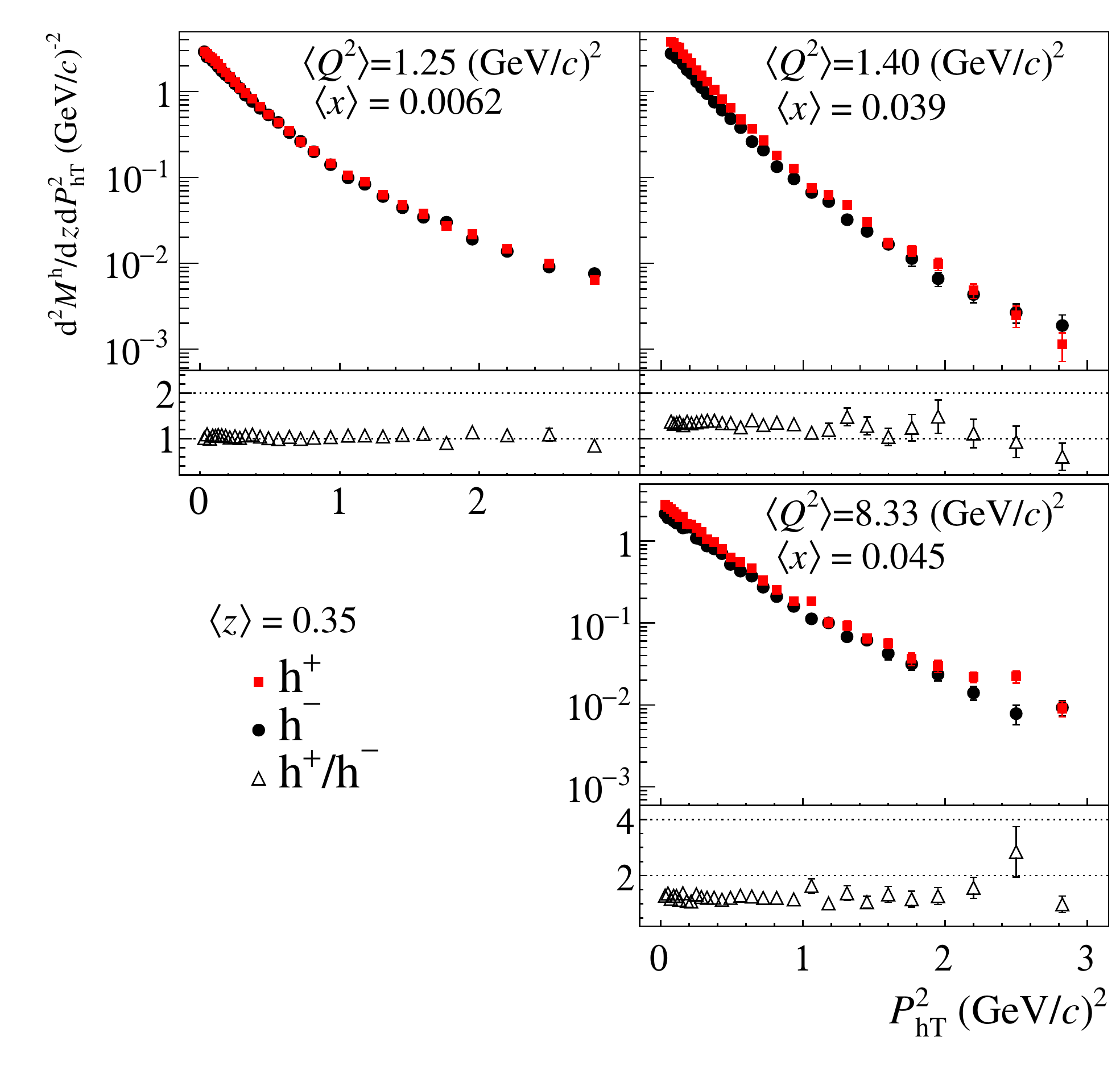}
\caption{Top row: Upper panels: Multiplicities of positively (full squares) and negatively (full circles) charged hadrons as a function of $P_{\rm{hT}}^{2}$ at fixed $Q^2$, i.e. $\langle Q^2\rangle \simeq 1.3$~(GeV/$c$)$^2$, for lower (left) and higher (right) $x$ bins. Lower panels: Ratio of multiplicities of positively and negatively charged hadron. Right column: Same at fixed $x$, i.e. $\langle x \rangle \simeq 0.04$, for lower (top) and higher (bottom) $Q^2$ bins. All measured at $\langle z\rangle = 0.35$. Only statistical uncertainties are shown.}
\label{mulfixedz}
\end{figure}

\begin{figure}[!h]
\centering
\includegraphics[height=10.2cm,width=0.6\textwidth]{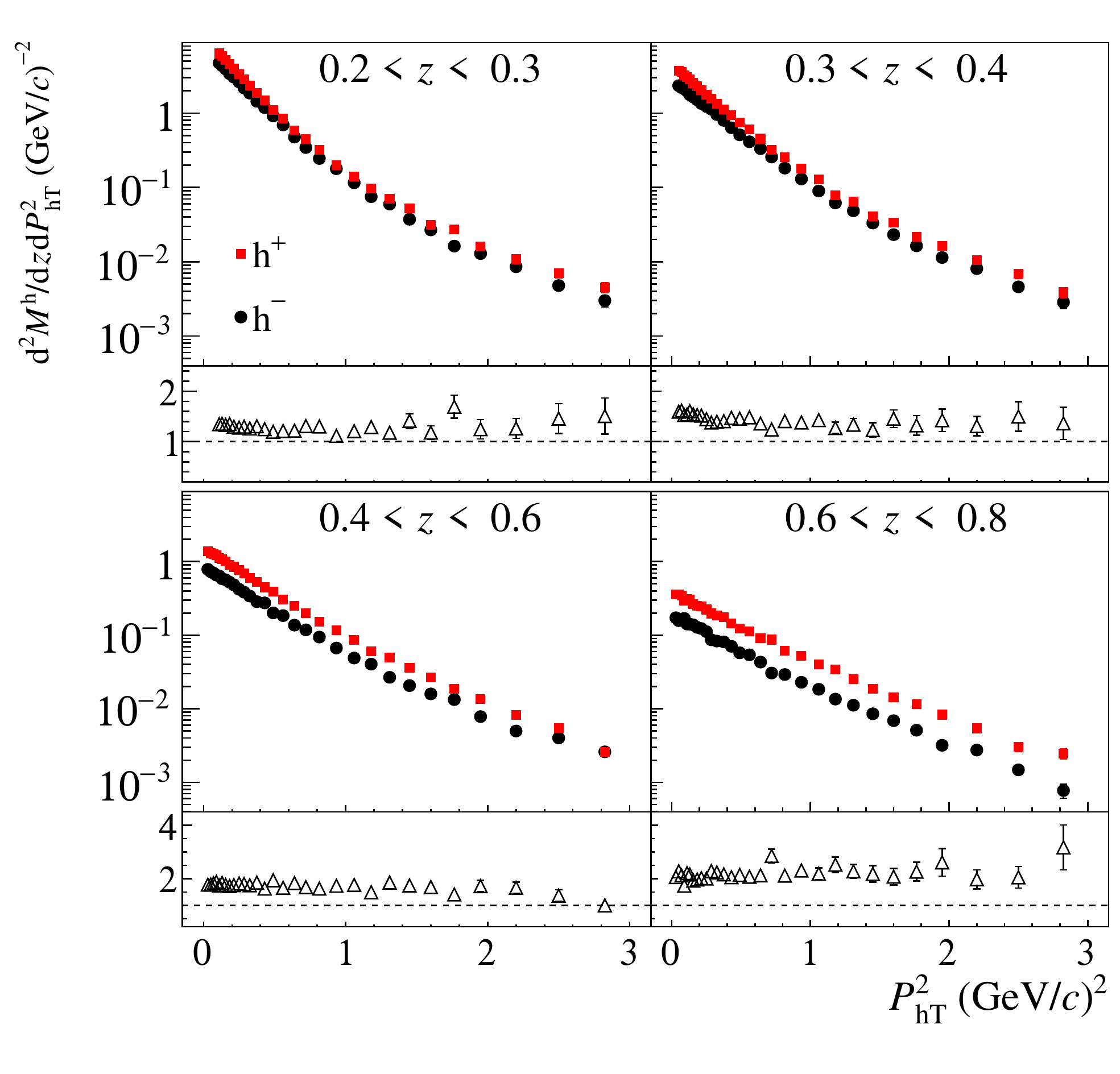}
\caption{Upper panels: Multiplicities of positively (full squares) and negatively (full circles) charged hadrons as a function of $P_{\rm{hT}}^{2}$ in four $z$ bins at $\langle Q^{2} \rangle = 9.78$~(GeV/$c$)$^2$ and $\langle x \rangle=0.149$. Lower panels: Ratio of multiplicities of positively and negatively charged hadron. Only statistical uncertainties are shown.} 
\label{mz}
\end{figure}

Multiplicities are larger for positively than for negatively charged hadrons. This difference significantly increases as $x$ increases and shows a weak variation with $Q^2$. It is observed to also depend on $z$ and it increases in the range of large $z$, i.e. $z > 0.4$, which confirms the observations made in Ref.~\cite{Adolph:2016bga}. Besides their magnitude, the $P_{\rm{hT}}$ dependence of the multiplicities shows a significant variation with $x$ at fixed $Q^{2}$ (as well as with $Q^2$ at fixed $x$) for any interval of $z$. These observations are separately illustrated in Figs.~\ref{mulfixedz} and \ref{mz} and discussed in detail in the following.

The comparison between the multiplicities of positively and negatively charged hadron is illustrated as a function of $x$ and $Q^2$ in Fig.~\ref{mulfixedz} for $\langle z\rangle = 0.35$. On the top row, $\rm{h}^{+}$ and $\rm{h}^{-}$ multiplicities are presented at $\langle Q^2\rangle \simeq 1.3$~(GeV/$c$)$^2$ in the smallest and the largest $x$ bins with average values $\langle x\rangle=0.0062$ and $\langle x\rangle=0.039$, respectively. In the right column, $\rm{h}^{+}$ and $\rm{h}^{-}$ multiplicities are similarly presented at $\langle x\rangle \simeq 0.04$ in the smallest and the largest $Q^2$ bins with average values $1.4$~(GeV/$c$)$^2$ and $8.3$~(GeV/$c$)$^2$, respectively. 
At fixed $Q^2$, the ratio of $\rm{h}^{+}$ to $\rm{h}^{-}$ multiplicities ranges from about 1 in the first $x$ bin to about 1.3 in the last $x$ bin. This increase as a function of $x$ confirms the expectation from valence u-quark dominance, i.e. the dominance of scattering off $u$-quarks. 
At fixed $x$, the ratio of $\rm{h}^{+}$ to $\rm{h}^{-}$ multiplicities decreases from 1.3 in the first $Q^2$ bin to about 1.2 in the last $Q^2$ bin.
While no significant difference is observed in the $P_{\rm{hT}}^{2}$-dependence of $\rm{h}^{+}$ and $\rm{h}^{-}$ multiplicities, the $P_{\rm{hT}}^{2}$-dependence of the multiplicities is observed to flatten at large values of $P_{\rm{hT}}^{2}$, where contributions from higher-order QCD processes like QCD Compton and photon-gluon fusion (PGF) are expected to dominate. The data suggest that flattening occurs both as $Q^{2}$ increases (at fixed $x$) and when $x$ decreases (at fixed $Q^{2}$).

In Figure~\ref{mz}, the comparison between $\rm{h}^{+}$ and $\rm{h}^{-}$ multiplicities is illustrated as a function of $z$. The multiplicities are presented as a function of $P_{\rm{hT}}^{2}$ in the four $z$ intervals in a given ($x$,~$Q^{2}$) bin with average values $\langle x \rangle=0.149$ and $\langle Q^{2} \rangle = 9.78$~(GeV/$c$)$^2$. A high $x$ bin is chosen, where the difference in the magnitude of the multiplicities is most recognisable. The ratio of $\rm{h}^{+}$ to $\rm{h}^{-}$ ranges from about 1.1 in the first $z$ bin to about 2 in the last $z$ bin, reflecting the fact that part of the negative hadrons (K$^{-}$ and $\bar{\rm{p}}$) can not be produced by the favoured fragmentation of a nucleon valence quark, which enhances the expected flavour dependence of TMD-FFs.
Another feature of the data is the variation of the $P_{\rm{hT}}^{2}$-dependence with increasing $z$ for both  small and large values of $P_{\rm{hT}}^{2}$. In particular, the data show a tendency to flatten at large $P_{\rm{hT}}$ as $z$ decreases, which emphases a significant $z$-dependence of the hadron transverse momentum with respect to the transverse momentum of the fragmenting quark, $p_{\perp}$.

\begin{figure}[!h]
\centering
\subfigure[]{\label{smallpTlargez_1}\includegraphics[height=7.5cm,width=.48\textwidth]{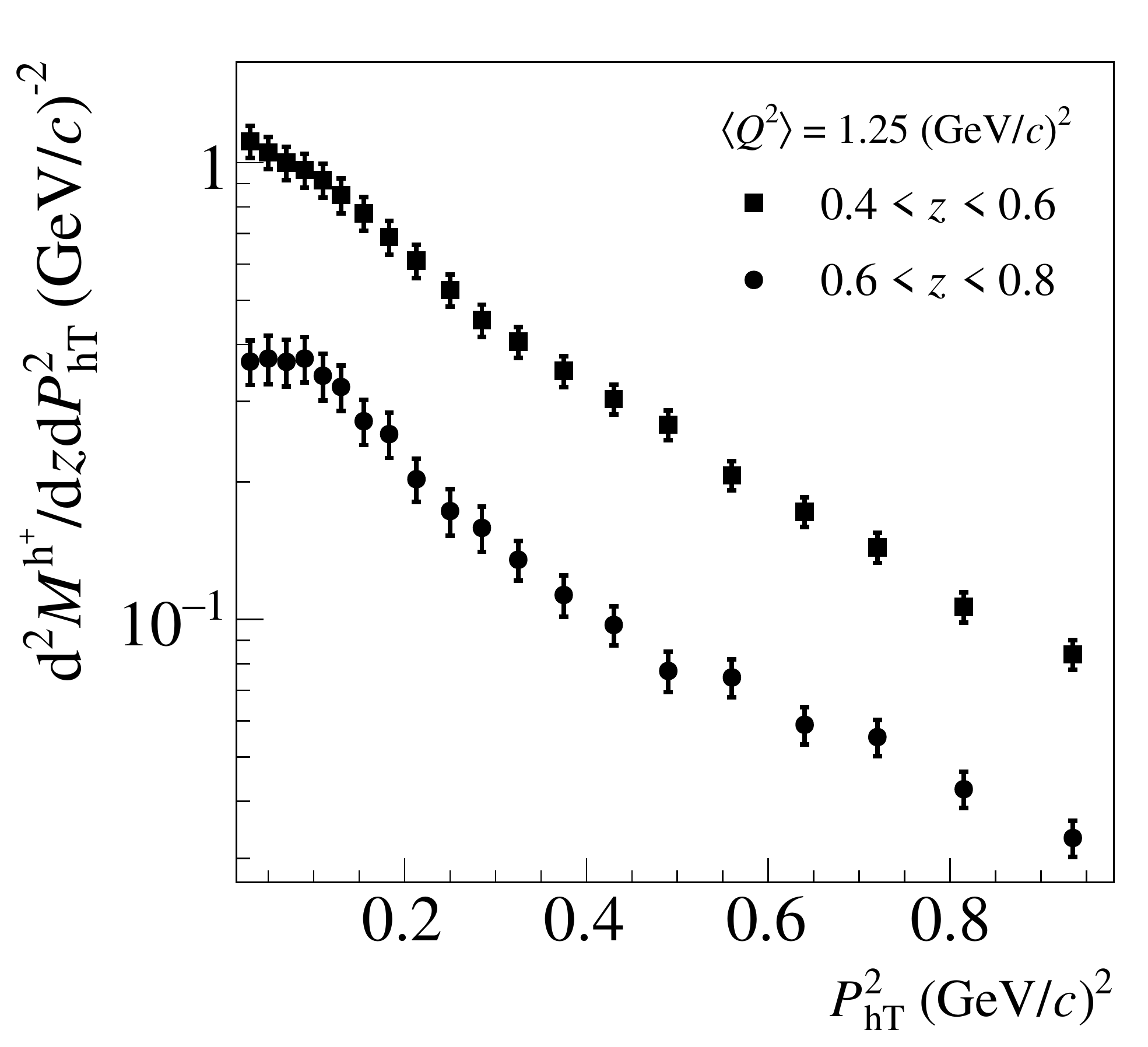}}
\subfigure[]{\label{smallpTlargez_2}\includegraphics[height=7.5cm,width=.48\textwidth]{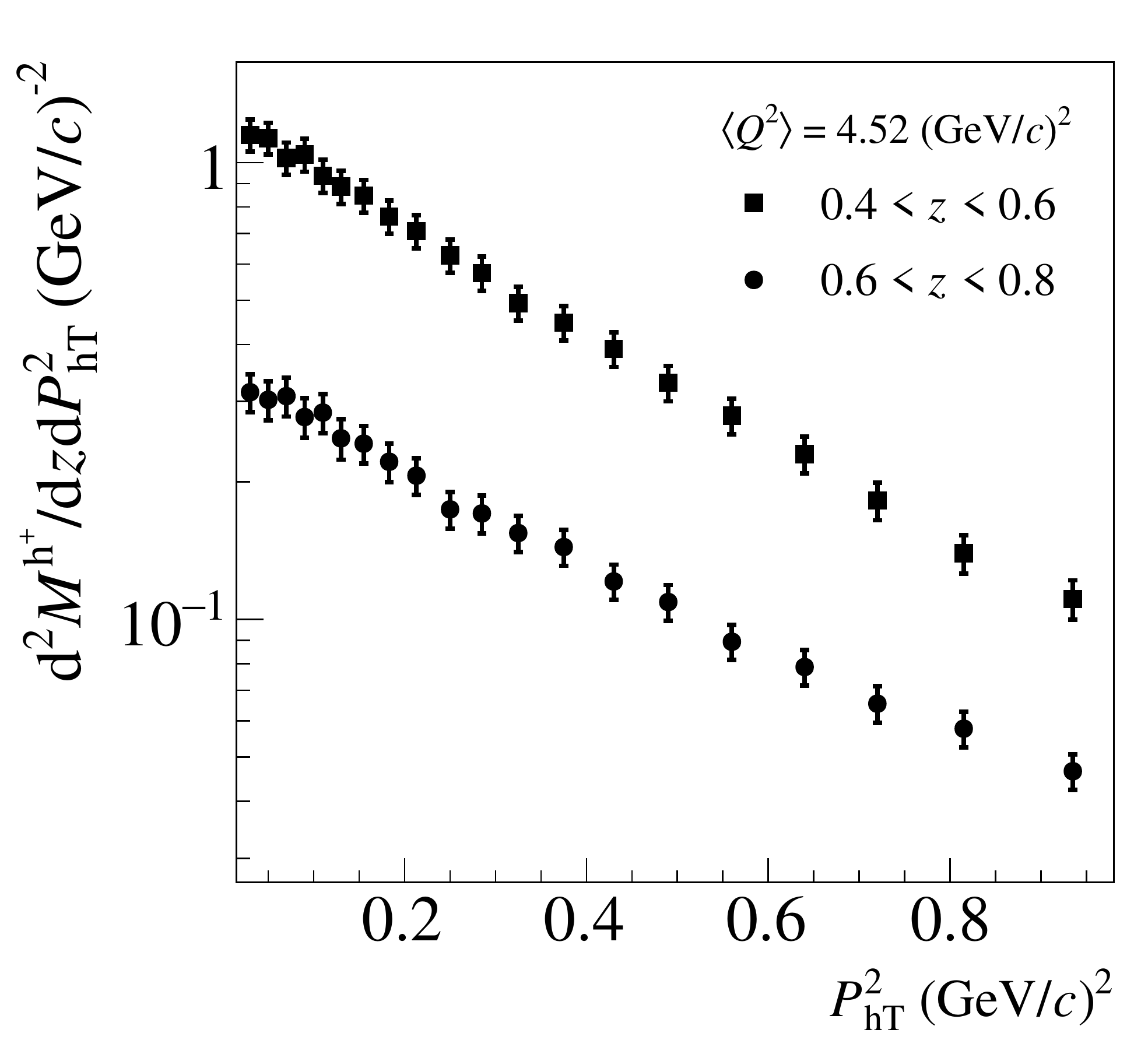}}
\caption{ (a) Multiplicities of positively charged hadrons as a function of $P_{\rm{hT}}^{2}$  at $\langle Q^2\rangle=1.25$~(GeV/$c$)$^2$ and $\langle x\rangle=0.006$ in two $z$ bins: $0.4 < z < 0.6$ and $0.6 < z < 0.8$. (b) Same as (a) at $\langle Q^2\rangle=4.52$~(GeV/$c$)$^2$ and $\langle x\rangle=0.043$. Statistical and systematic uncertainties are added in quadrature and shown.}
\label{smallpTlargez}
\end{figure}

Another intriguing effect is observed in the kinematic domain 1~(GeV/$c$)$^2$ $< Q^2 < 1.7$~(GeV/$c$)$^2$ and $0.6 < z < 0.8$, in the range of small $P_{\rm{hT}}^{2}$. Charged hadron multiplicities do not exhibit an exponential form in $P_{\rm{hT}}^{2}$ in this kinematic region and show an unexpected flat dependence at very small values of $P_{\rm{hT}}^{2}$. This effect is also present in the earlier published~\cite{Adolph:2013stb} distributions of charged hadrons as a function of $P_{\rm{hT}}^{2}$. It is illustrated in Fig.~\ref{smallpTlargez}, which shows the multiplicity of positive hadrons as a function of $P_{\rm{hT}}^{2}$ up to 0.8~(GeV/$c$)$^2$ at $\langle Q^{2}\rangle = 1.25$~(GeV/$c$)$^{2}$ and $\langle x\rangle = 0.0062$ (left-hand side) and at $\langle Q^{2}\rangle = 4.52$~(GeV/$c$)$^{2}$ and $\langle x\rangle = 0.043$ (right-hand side). It should be noted that this particular kinematic region suffers from the highest contribution of the $\rho^{0}$ decay products to the charged hadron sample (Fig.~\ref{C_SIDIS}, blue curve) evaluated using the MC simulation. This effect is further discussed in Sec.~\ref{fitsection}.

The multiplicities shown in Figs \ref{fullmul0}-\ref{smallpTlargez} agree with the previous measurement of hadron distributions performed by COMPASS~\cite{Adolph:2013stb}. However, as mentioned in Sec.~\ref{intro}, this measurement considerably extends the kinematic range and reduces the statistical and systematic uncertainties, in particular the uncertainties on the normalisation of the $P_{\rm{hT}}^{2}$-integrated multiplicities. 

\clearpage

%~~~~~~~~~~~~~~~~~~~~~~~~~~~~~~%
%                                                               %
%                  subsection n.2                      %
%                                                               %
%~~~~~~~~~~~~~~~~~~~~~~~~~~~~~~%

\subsection{Comparison with other measurements}
\label{results2}
\vspace{0.3cm}

\begin{figure}[htp]
\centering
\includegraphics[height=7.cm,width=1.1\textwidth]{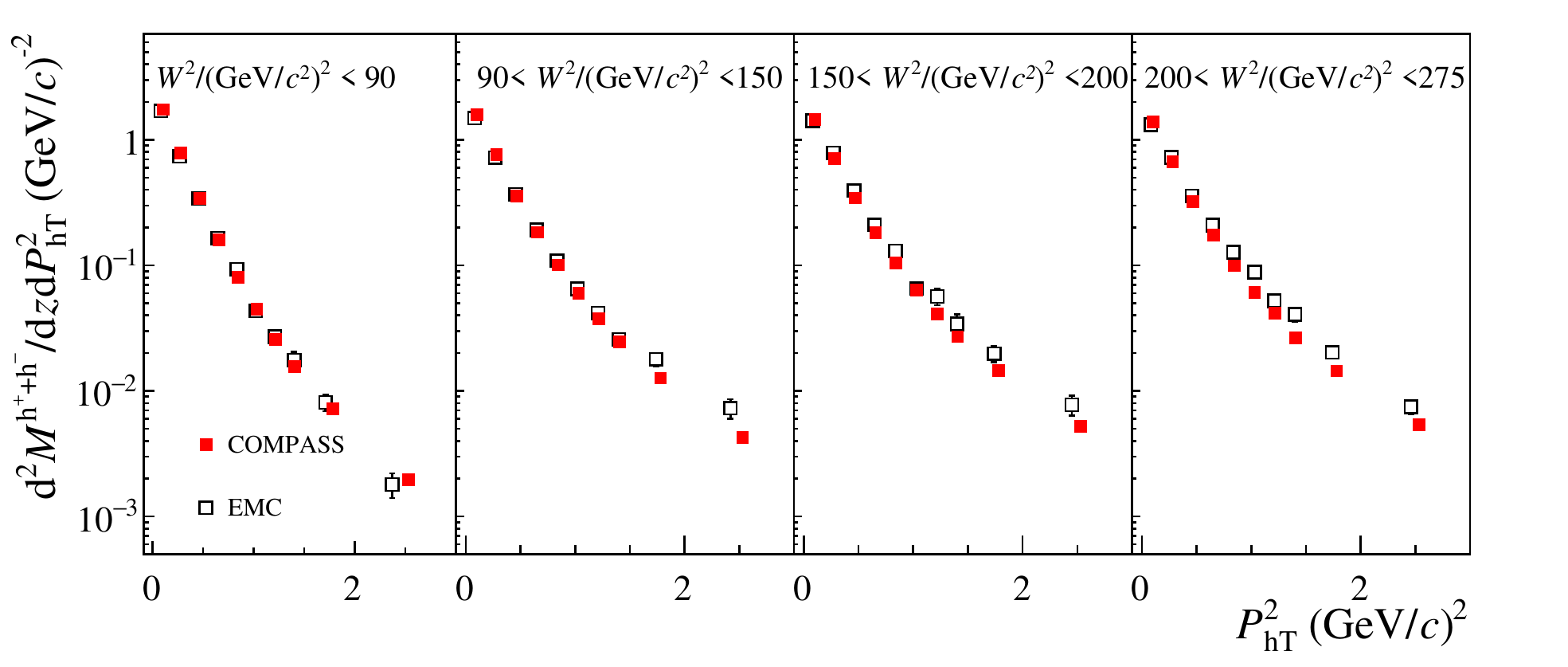}
\caption{Charged hadron multiplicities from COMPASS (beam energy 160~GeV) compared to EMC results (beam energies 100~GeV to 280~GeV)~\cite{Ashman:1991cj}, shown in four bins of $W^2$, which have the following mean values in (GeV/$c^2$)$^2\colon$ 59.4, 113.8, 174.3 and 236. Only statistical uncertainties are shown.}
\label{CMPvsEMC1}
\end{figure}

\begin{figure}[htp]
\centering
\includegraphics[height=11.cm,width=.7\textwidth]{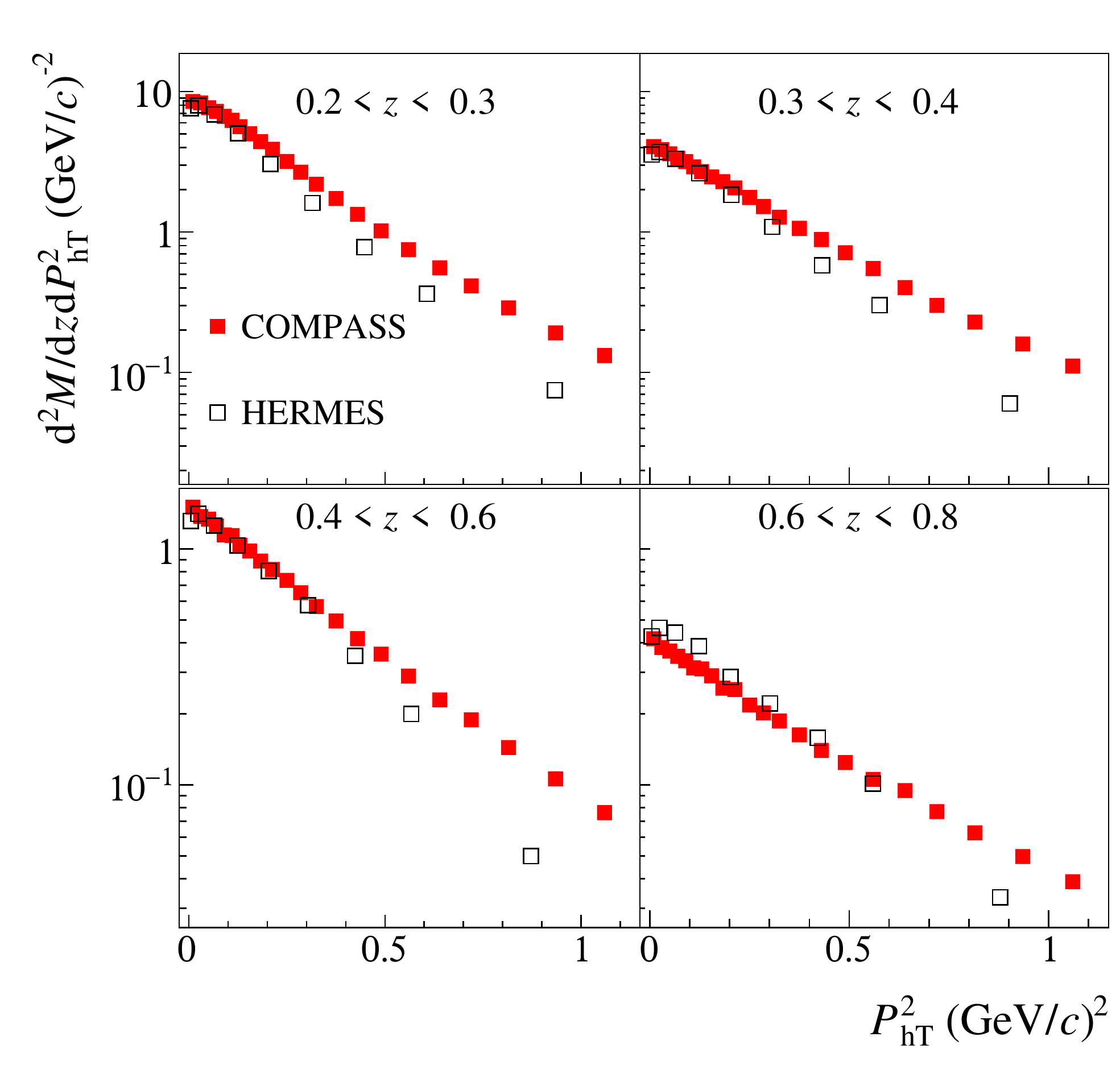}
\caption{Multiplicities of positively charged hadrons from COMPASS (beam energy 160~GeV), compared in four $z$ bins to the results of HERMES (beam energy 27.6~GeV) on positively charged pions~\cite{Airapetian:2012ki}. The multiplicities from COMPASS are obtained by integrating the present results over $x$ and $Q^2$. Only statistical uncertainties are shown.}
\label{CMPvsHMS}
\end{figure}

\begin{figure}[htp]
\centering
\includegraphics[height=8.cm,width=.45\textwidth]{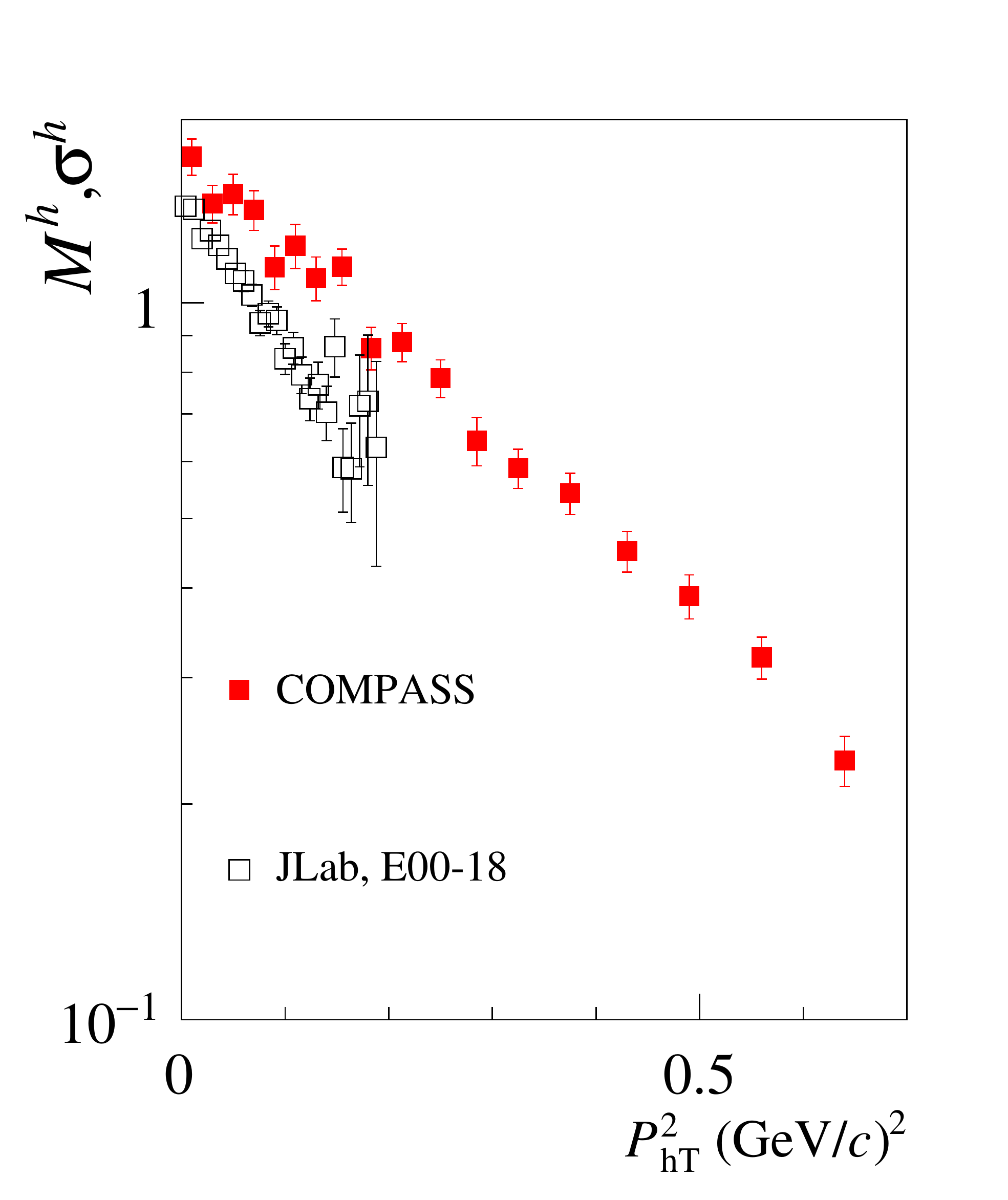}
\caption{Multiplicities $M^h$ of positively charged hadrons from COMPASS (beam energy 160~GeV), compared to the cross section $\sigma^{h}$ for positively charged pions as measured by experiment E00-18 (beam energy 5.479 GeV) at Jefferson Lab~\cite{Asaturyan:2011mq}. Both results are not corrected for diffractive vector-meson production. The $Q^2$ range is 2~(\rm{GeV}/$c$)$^2 < Q^2 < 4$~(\rm{GeV}/$c)^2$ for E00-18 and 7~(GeV/$c)^2 < Q^2 < 16$~(GeV/$c)^2$ for COMPASS. Only statistical uncertainties are shown.}
\label{CMPvsJLAB}
\end{figure}

The multiplicities presented above are compared in Figs.~\ref{CMPvsEMC1}-\ref{CMPvsJLAB} to results from previous semi-inclusive measurements in similar kinematic regions. The experiments are compared in Table~\ref{ExistingMeas}.

\begin{table}[htp]
\hspace{-1cm}
\caption{Comparison of the main features of experiments that performed semi-inclusive measurements in deep inelastic scattering. The subscript (min) in $Q^2$ and $W^2$ refers to the lower limit.}
\begin{center}
\begin{tabular}{c|c|c|c|c|p{1.9cm}|} \cline{2-6}
								&	EMC~\cite{Ashman:1991cj} 
								&	HERMES~\cite{Airapetian:2012ki}				
								&	JLAB~\cite{Asaturyan:2011mq}		
								&	COMPASS~\cite{Adolph:2013stb}	
								& COMPASS (This paper) \\ \hline
\multicolumn{1}{ |l| }{Target 			}&	p/d			&	p/d					& 	d			&	d				&	d		\\
\multicolumn{1}{ |l| }{Beam energy (GeV)	}&	100-280	&	27.6					&	5.479		&	160				&	160		\\
\multicolumn{1}{ |l| }{Hadron type		}&	$h^{\pm}$			&	$\pi^{\pm}$,K$^{\pm}$	&	$\pi^{\pm}$	&	$h^{\pm}$			&	$h^{\pm}$	\\		
\multicolumn{1}{ |l| }{Observable		}&	$M^{h^{+}+h^{-}}$	&	$M^{h}$				&	$\sigma^{h}$	&	$M^{h}$			&	$M^{h}$	\\
\multicolumn{1}{ |l| }{$Q^2_{\rm{min}}$ (GeV/$c$)$^2$}& 2/3/4/5		& 	1					&	2			&	1				&	1		\\
\multicolumn{1}{ |l| }{$W^2_{\rm{min}}$ (GeV/$c^2$)$^2$ }& -				& 	10					&	4			&	25				& 	25	\\
\multicolumn{1}{ |l| }{$y$ range			}&	[0.2,0.8]			& 	[0.1,0.85]				&	[0.1,0.9]		& 	[0.1,0.9]			&	[0.1,0.9]	\\
\multicolumn{1}{ |l| }{$x$ range			}& 	[0.01,1]			&	[0.023,0.6]			& 	[0.2,0.6]		&	[0.004,0.12]		&	[0.003,0.4] \\
\multicolumn{1}{ |l| }{$P_{\rm{hT}}^{2}$ range (GeV/$c$)$^2$ }& [0.081,	15.8] &[0.0047,0.9]  & [0.004,0.196] & [0.02,0.72] & [0.02,3] \\
\hline
\end{tabular}
\end{center}
\label{ExistingMeas}
\end{table}%

In order to compare the present COMPASS results on TMD hadron multiplicities with the corresponding ones by EMC~\cite{Ashman:1991cj}, our data sample is reanalysed in bins of $z$ and $W^2$ according to the binning given in Ref.~\cite{Ashman:1991cj}. The EMC measurements are performed in slightly different kinematic ranges in $Q^2$ and $y$, as shown in Tab.~\ref{ExistingMeas}. While for the measurement described in this paper a deuteron target was used, EMC used proton and deuteron targets and also four different beam energies, which led to four different kinematic ranges. The comparison shown in Fig.~\ref{CMPvsEMC1}, where the sum of $\rm{h}^{+}$ and $\rm{h}^{-}$ multiplicities is presented as a function of $P_{\rm{hT}}^{2}$ in four $W^2$ bins in the range $0.2 < z < 0.4$, demonstrates good agreement between COMPASS and EMC results. According to the study in~Ref.~\cite{Anselmino:2006rv}, the $P_{\rm{hT}}^{2}$-dependence of the EMC data could be explained in the simple collinear parton model up to 8 (GeV/$c$)$^2$ in $P_{\rm{hT}}^{2}$.

In Figure~\ref{CMPvsHMS}, the multiplicities of positively charged hadrons are compared in the four bins of $z$ to the multiplicities of positively charged pions measured by the HERMES Collaboration~\cite{Airapetian:2012ki}, where both were corrected for diffractive vector-meson contribution. The measurements by HERMES cover the kinematic range $Q^2 > 1$ (GeV/$c$)$^2$ and $0.023 < x < 0.6$. For this comparison, the COMPASS h$^{+}$ multiplicities are integrated over $x$ in the closest possible range $0.02 < x < 0.4$ and also over $Q^2$. 
It should be noted that the two experiments cover different ranges in $Q^2$. While the highest $Q^2$ value reached by HERMES is $15$ (GeV/$c$)$^2$, COMPASS reaches $81$ (GeV/$c$)$^2$.
Despite this difference, a reasonable agreement in the magnitude of the measured multiplicities is found for $z < 0.6$ and small $P_{\rm{hT}}^{2}$. Most likely due to the differences in kinematic coverage, the agreement between the two sets is rather modest, and the data sets exhibit different dependences upon $P_{\rm{hT}}^{2}$. In addition, a dip is observed in the HERMES data at very small transverse momenta, i.e. $P_{\rm{hT}}^{2}\sim 0.05$ (GeV/$c$)$^2$. This dip, which is not observed in the shown $Q^2$-integrated distribution, appears to be very similar to the trend shown in Fig.~\ref{smallpTlargez} by the COMPASS data at low $Q^{2}$.

In Figure~\ref{CMPvsJLAB}, the h$^{+}$ multiplicities are compared to the $\pi^{+}$ semi-inclusive cross section measured by the E00-18 experiment~\cite{Asaturyan:2011mq} at Jefferson Lab. The measurement by the E00-18 was performed at $\langle z \rangle=0.55$ and $\langle x\rangle=0.32$ in the range 2~(GeV/$c$)$^2 < Q^2 <$ 4~(GeV/$c$)$^2$. The COMPASS results are given at similar ($x$,~$z$) values, i.e. $\langle z\rangle = 0.5$, $\langle x\rangle= 0.3$, and span the range  7 (GeV/$c)^2$ $< Q^2 <$16 (GeV/$c)^2$. Similar to the case of the comparison of COMPASS and HERMES data shown in Fig.~\ref{CMPvsHMS}, here the observed different $P_{\rm{hT}}$-dependence could be due to the different $Q^2$ values of the two measurements.

% 	COMPASS vs EMC

%	COMPASS vs HERMES

%	COMPASS vs JLab

\clearpage
%~~~~~~~~~~~~~~~~~~~~~~~~~~~~~~~~~~~~~~~~~~~~~~%
%                                                                                                 %
%                   New section                                                          %
%                                                                                                 %
%~~~~~~~~~~~~~~~~~~~~~~~~~~~~~~~~~~~~~~~~~~~~~~%
\section{\large{Fits of the measured hadron multiplicities}}
\label{fitsection}
\vspace{0.3cm}

%~~~~~~~~~~~~~~~~~~~~~~~~~~~~~~%
%                                                               %
%                  subsection n.1                      %
%                                                               %
%~~~~~~~~~~~~~~~~~~~~~~~~~~~~~~%

\subsection{The range of small $P_{\rm{hT}}$}
\vspace{0.3cm}

The $P_{\rm{hT}}$-dependence of the cross section for semi-inclusive measurements of hadron leptoproduction was empirically reasonably well described by a Gaussian parameterisation for the $k_{\rm{T}}$- and $p_{\rm{h}\perp}$-~dependence of TMD-PDFs and TMD-FFs in the range of small $P_{\rm{hT}}$, i.e. $P_{\rm{hT}} < 1$~(GeV/$c$). This Gaussian parameterisation leads to a $P_{\rm{hT}}^{2}$-dependence of the multiplicities of the form$\colon$ 

\begin{equation}
\frac{\dd^{2} M^{\rm{h}}(x,Q^2;z)}{\dd z\dd P_{\rm{hT}}^{2}} = \frac{N}{\langle P_{\rm{hT}}^{2}\rangle}\exp \left( -\frac{P_{\rm{hT}}^{2}}{\langle P_{\rm{hT}}^{2}\rangle} \right),
\label{1expfunction}
\end{equation}

where the normalisation coefficient $N$ and the average transverse momentum $\langle P_{\rm{hT}}^{2}\rangle$, i.e. the absolute value of the inverse slope of the exponent in Eq.~\ref{1expfunction}, are functions of $x$, $Q^2$ and $z$.

A fairly good description of SIDIS~\cite{Anselmino:2005nn} data was reached with the Gaussian parameterisation without considering neither the $z$ nor the quark flavour dependence of TMD-FFs.
Recent semi-inclusive measurements of transverse-momentum-dependent hadron multiplicities~\cite{Airapetian:2012ki} and distributions~\cite{Adolph:2013stb} aimed at an extraction of both $\langle k_{\perp}^{2}\rangle$ and $\langle p_{\perp}^{2}\rangle$. These two observables, however, were found to be too strongly anti-correlated to be disentangled~\cite{Anselmino:2013lza,Signori:2013mda,radici}. In order to extract them, a combined analysis of both the differential transverse-momentum-dependent hadron multiplicities and the spin-independent azimuthal asymmetries in SIDIS may be required.
In the following we will discuss separately fits in the region of small $P_{\rm{hT}}$ and in the full range of $P_{\rm{hT}}^{2}$ accessible by COMPASS, i.e. 0.02~(GeV/$c)^2 < P_{\rm{hT}}^2 <$ 3~(GeV/$c)^2$.

\begin{figure}[htp]
\centering
\includegraphics[height=8.cm,width=.9\textwidth]{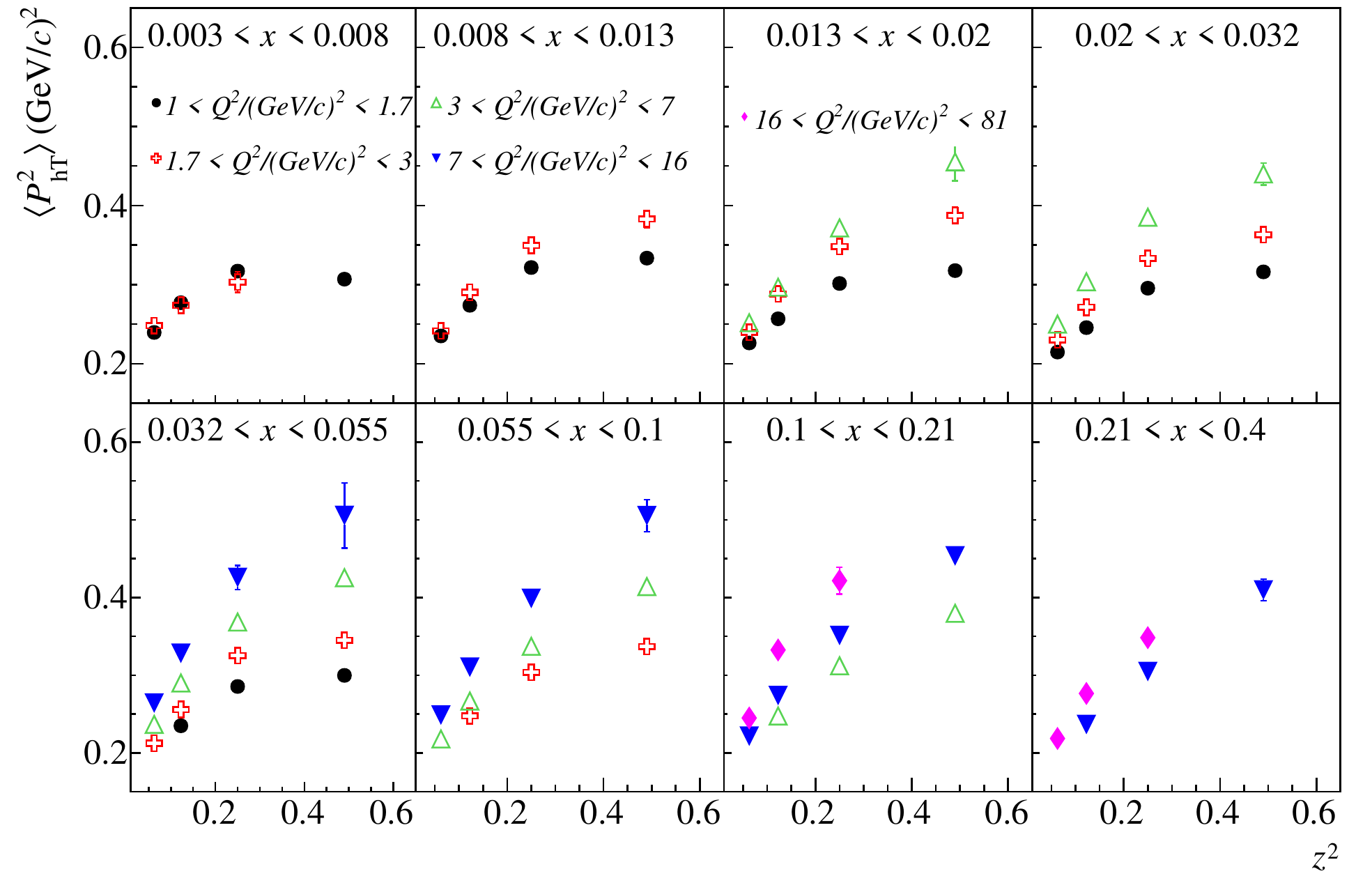}
\caption{Average transverse momentum $\langle P_{\rm{hT}}^{2}\rangle$, as obtained from the fit of h$^{+}$  multiplicities using the single Gaussian parameterisation, shown as a function of $z^{2}$. The eight panels correspond to the eight $x$-bins as indicated, where in each panel data points from all five $Q^2$ bins are shown. Error bars denote to statistical uncertainties.}
%: [1,1.7] (black symbols), [1.7,3] (red symbols), [3,7] (green symbols), [7,16] (blue symbols) and [16,81] (violet symbols). Error bars denote to statistical uncertainties.}
\label{1exp}
\end{figure}

The hadron multiplicities presented in Figs.~\ref{fullmul0}-\ref{fullmul3} are fitted in each ($x$, $Q^2$, $z$) kinematic bin in the range 0.02~(GeV/$c$)$^2<P_{\rm{hT}}^{2}< 0.72$~(GeV/$c$)$^2$ using the single-exponential function given in Eq.~\ref{1expfunction}.
Using only statistical uncertainties in the fit, reasonable values of $\chi^{2}$ per degree of freedom ($\chi^{2}_{\rm{dof}}$) are obtained in all ($x$,~$Q^2$) bins, except for low values of $Q^2$ and small values of $z$, i.e. $z < 0.3$, where the $\chi^2_{\rm{dof}}$ values are significantly larger than 3 in most of the $x$ bins. Including the systematic uncertainties in the fit by adding them in quadrature to the statistical ones significantly improves the values of $\chi^2_{\rm{dof}}$, whereas the fitted parameters remain unchanged. 
 The $z^{2}$-dependence of $\langle P_{\rm{hT}}^{2}\rangle$ obtained from the fits is shown in Fig.~\ref{1exp} for $\rm{h}^{+}$ in the five $Q^2$ bins available in a given $x$ bin. A non-linear dependence of $\langle P_{\rm{hT}}^{2}\rangle$ on $z^2$ is observed in the range of small $x$ and $Q^{2}$, in contrast to the range of large $x$ and $Q^2$ where it becomes linear. In addition, $\langle P_{\rm{hT}}^{2}\rangle$ significantly increases with $Q^2$ at fixed $x$ and $z$, especially at high $z$. The h$^{+}$ multiplicities have larger values of $\langle P_{\rm{hT}}^{2}\rangle$ than the h$^{-}$ ones at large $z$, while no significant difference is observed at small $z$. This conclusion confirms the one made in our previous publication~\cite{Adolph:2013stb}, where a detailed study of the kinematic dependence of $\langle P_{\rm{hT}}^{2}\rangle$ was presented and discussed.

\begin{figure}[!h]
\centering
\subfigure[]{\label{SMALLPTLARGEZ1}\includegraphics[height=8.3cm,width=.4\textwidth]{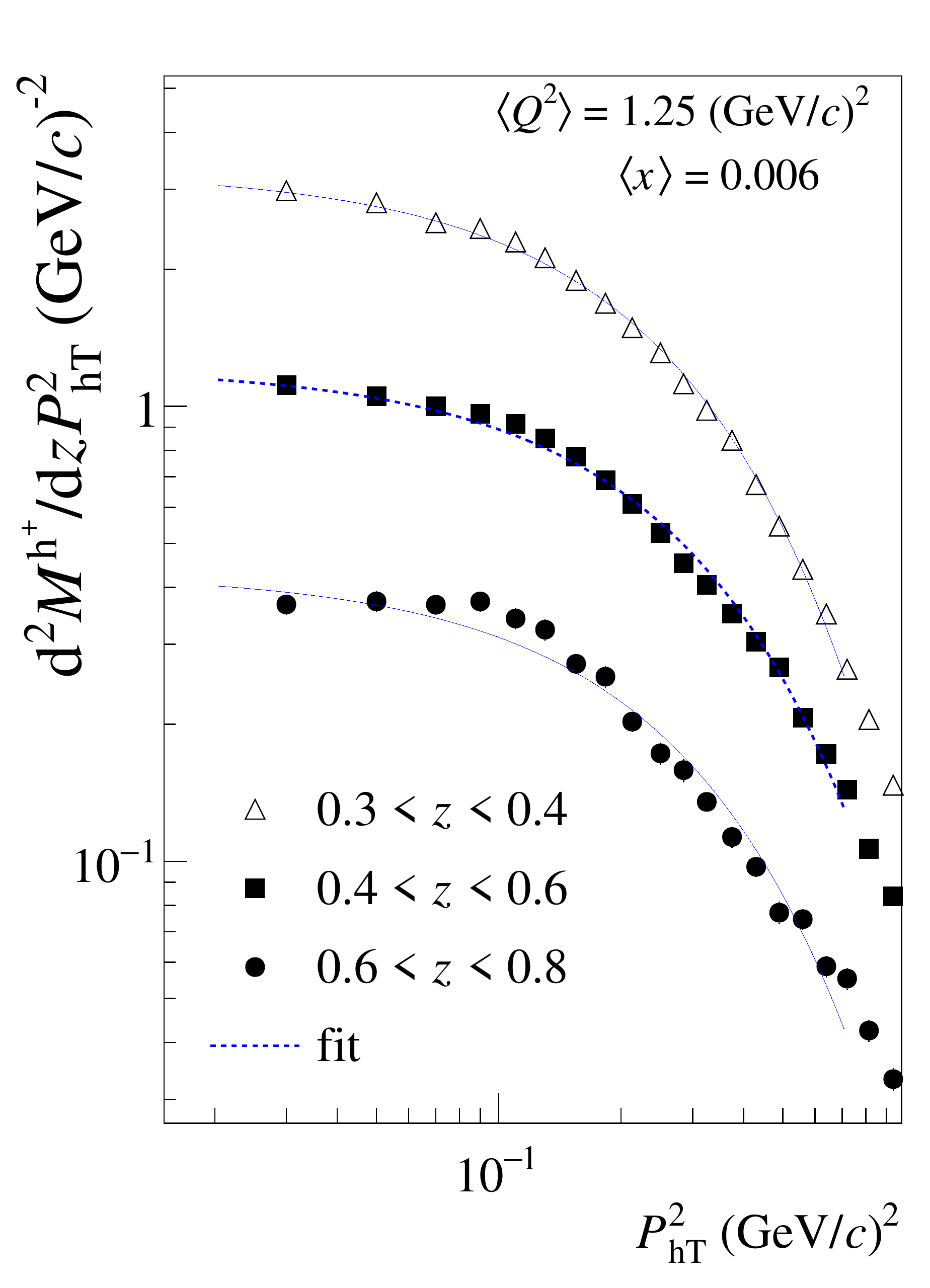}} 
\subfigure[]{\label{SMALLPTLARGEZ2}\includegraphics[height=8.3cm,width=.4\textwidth]{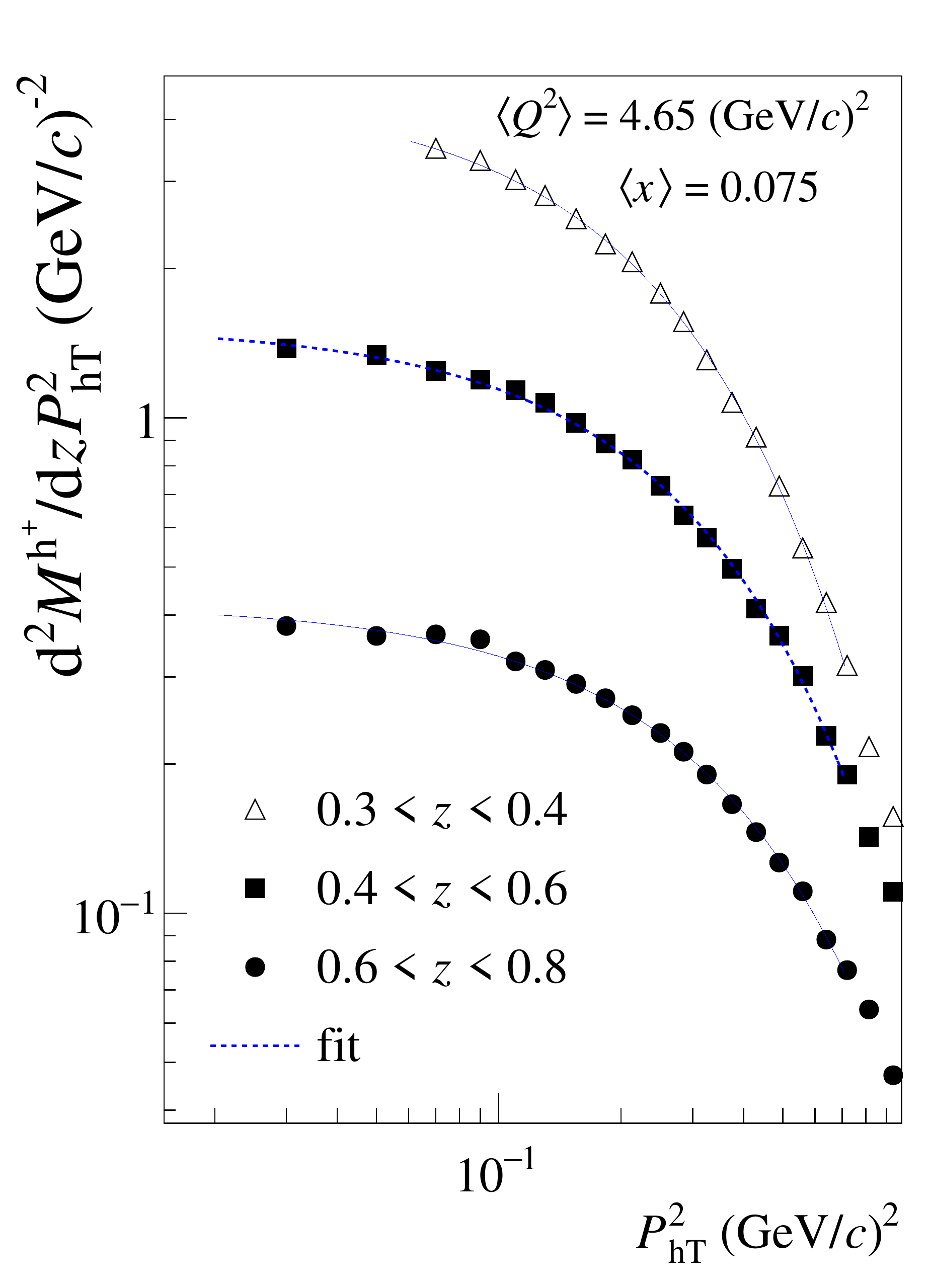}} 
\caption{(a) h$^{+}$ multiplicities as a function of $P_{\rm{hT}}^2$ up to 1~(GeV/$c$)$^2$ for three $z$ bins at $\langle Q^2\rangle=1.25$~(GeV/$c$)$^2$ and $\langle x\rangle=0.006$. The curves correspond to the fits using Eq.~\ref{1expfunction}. (b) Same as (a) for  $\langle Q^2\rangle=4.65$~(GeV/$c$)$^2$ and $\langle x\rangle=0.075$.}
\label{SMALLPTLARGEZ}
\end{figure}

As mentioned earlier in Sec.~\ref{results2}, the kinematic region of small $Q^{2}$ and large $z$, i.e. $Q^2 < 1.7$~(GeV/$c$)$^2$ and $0.6 < z < 0.8$, shows an intriguing effect in the range of small $P_{\rm{hT}}^{2}$. As can be seen from Fig.~\ref{smallpTlargez}, in this range $\rm{h}^{+}$ and $\rm{h}^{-}$ multiplicities do not exhibit an exponential form in $P_{\rm{hT}}^{2}$ and show an unexpected flat dependence at very small values of $P_{\rm{hT}}^{2}$. Figure~\ref{SMALLPTLARGEZ1} shows the multiplicity of positively charged hadrons as a function of $P_{\rm{hT}}^{2}$ up to $0.8$~(GeV/$c$)$^2$ at $\langle Q^2\rangle=1.25$~(GeV/$c$)$^2$ and $\langle x\rangle=0.006$. While a single-exponential function reasonably describes the $P_{\rm{hT}}^{2}$-dependence for $0.3 < z < 0.4$, the experimental data clearly deviate from this functional form as $z$ increases, with $\chi^2_{\rm{dof}}$ values increasing from 1.8 in the smallest $z$ bin to 4.6 in the largest one. As an example, Fig.~\ref{SMALLPTLARGEZ2} shows h$^+$ multiplicities at larger $Q^2$, i.e. $\langle Q^2\rangle=4.65$~(GeV/$c$)$^2$ and $\langle x\rangle=0.075$, where the single-exponential function fits the data well in all $z$ bins.

The measured charged-hadron multiplicities show that in the range of small $P_{\rm{hT}}$, i.e. for $P_{\rm{hT}} < 1$~(GeV/$c)^2$, the simple parameterisation using a single-exponential function describes the $P_{\rm{hT}}^{2}$-dependence of the results quite well for not too large values of $Q^2$. For increasing $Q^2$, the $P_{\rm{hT}}^2$-dependence of the multiplicities changes as can be seen in Fig.~\ref{mulfixedz}. A more complex parameterisation appears to be necessary to fit the data, as shown in Ref.~\cite{Aidala:2014hva} .

%~~~~~~~~~~~~~~~~~~~~~~~~~~~~~~%
%                                                               %
%                  subsection n.2                      %
%                                                               %
%~~~~~~~~~~~~~~~~~~~~~~~~~~~~~~%

\subsection{The full measured $P_{\rm{hT}}$ range}
\vspace{0.3cm}

Up to now, only one study~\cite{Anselmino:2006rv} was performed to describe the full range in $P_{\rm{hT}}$ using a Gaussian parameterisation for the $k_{\rm{T}}$- and $p_{\rm{h}\perp}$- dependence of TMD-PDFs and TMD-FFs in the range $P_{\rm{hT}} < 1$~GeV/$c$, and calculating pQCD higher order collinear contributions in the range $P_{\rm{hT}} > 1$ GeV/$c$. A reasonable description of semi-inclusive hadron multiplicities and cross sections measured by the EMC~\cite{Ashman:1991cj} and ZEUS~\cite{Derrick:1995xg} Collaborations, respectively, was achieved. Below, we attempt to describe the $P_{\rm{hT}}^{2}$-dependence of the above presented charged-hadron multiplicities over the full $P_{\rm{hT}}$-range explored by COMPASS, i.e. 0.02~(GeV/$c)^{2} < P_{\rm{hT}}^2 < 3$~(GeV/$c)^{2}$, using the following two parameterisations$\colon$

\begin{equation}
\label{2expeq}
\quad F_{1} = \frac{N_{1}}{\alpha_{1}} \exp\left(-\frac{P_{\rm{hT}}^{2}}{\alpha_{1}}\right) + \frac{N_{1}^{\ensuremath{\prime}}}{\alpha_{1}^{\ensuremath{\prime}}}\exp\left(-\frac{P_{\rm{hT}}^{2}}{\alpha_{1}^{\ensuremath{\prime}}}\right),
\end{equation}

\begin{equation}
\quad F_{2} = N_{2} \left(1-(1-q)\frac{P_{\rm{hT}}^{2}}{T}\right)^{\frac{1}{1-q}}.
\label{tsalliseq}
\end{equation}

%~~~~~~~~~~~~~~~~~~~~~~~~~~~~~~~~~~~~~~~~~~~~%
%                                                                                            %
%			 Fit using 2 exponential 				   %
%                                                                                            % 
%~~~~~~~~~~~~~~~~~~~~~~~~~~~~~~~~~~~~~~~~~~~~%	

\begin{figure}[!h]
\centering
\includegraphics[height=9.5cm,width=.4\textwidth]{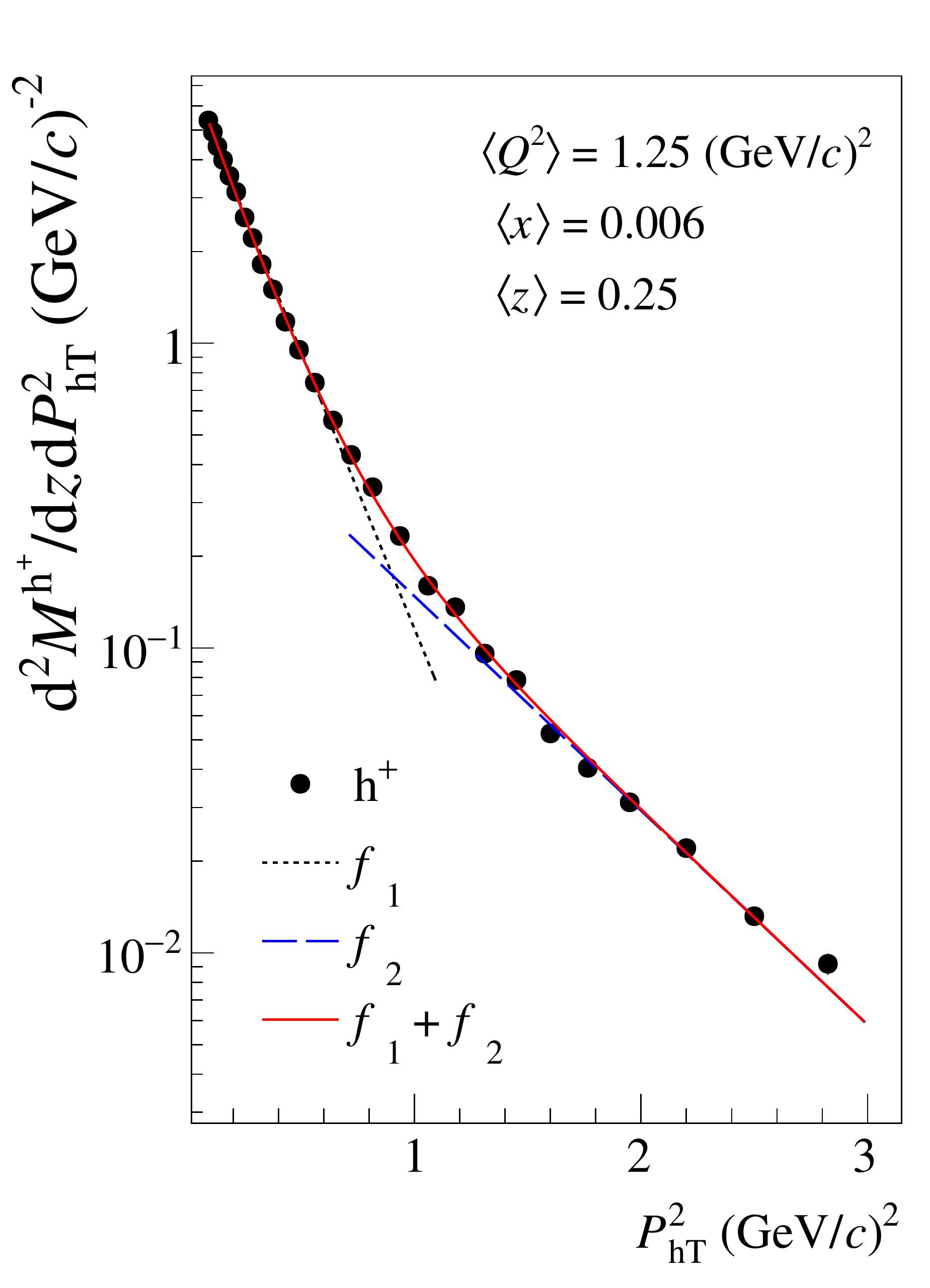}
\caption{Multiplicities of positively charged hadrons as a function of $P_{\rm{hT}}^{2}$ for $\langle Q^2 \rangle =1.25$~(GeV/$c$)$^{2}$, $\langle x\rangle=0.006$ and $\langle z\rangle=0.25$. The black dotted curve represents the first exponential function $f_1 = (N_{1}/\alpha_1)\exp(-P_{\rm{hT}}^{2}/\alpha_{1})$, the blue dashed curve represents the second exponential function $f_2 = (N_{1}^{\ensuremath{\prime}}/\alpha_1^{\ensuremath{\prime}})\exp(-P_{\rm{hT}}^{2}/\alpha_{1}^{\ensuremath{\prime}})$, and the red curve represents the sum ($f_1 + f_2$) as in Eq.~\ref{2expeq}. Only statistical uncertainties are shown and used in the fit.}
\label{muland12exp}
\end{figure}

\begin{figure}[!h]
\centering
\subfigure[]{\label{pos}\includegraphics[height=9.5cm,width=.46\textwidth]{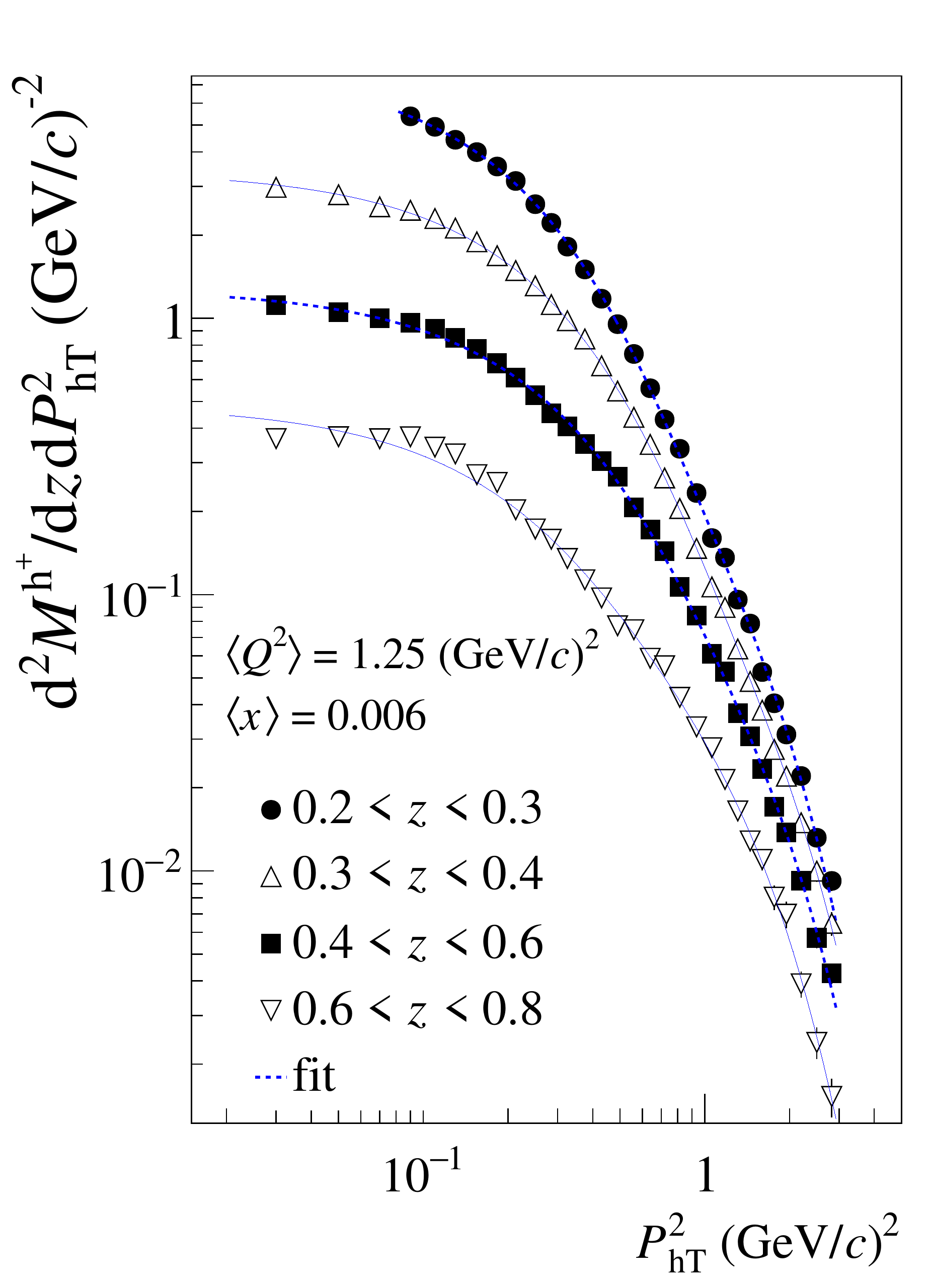}}
\subfigure[]{\label{neg}\includegraphics[height=9.5cm,width=.46\textwidth]{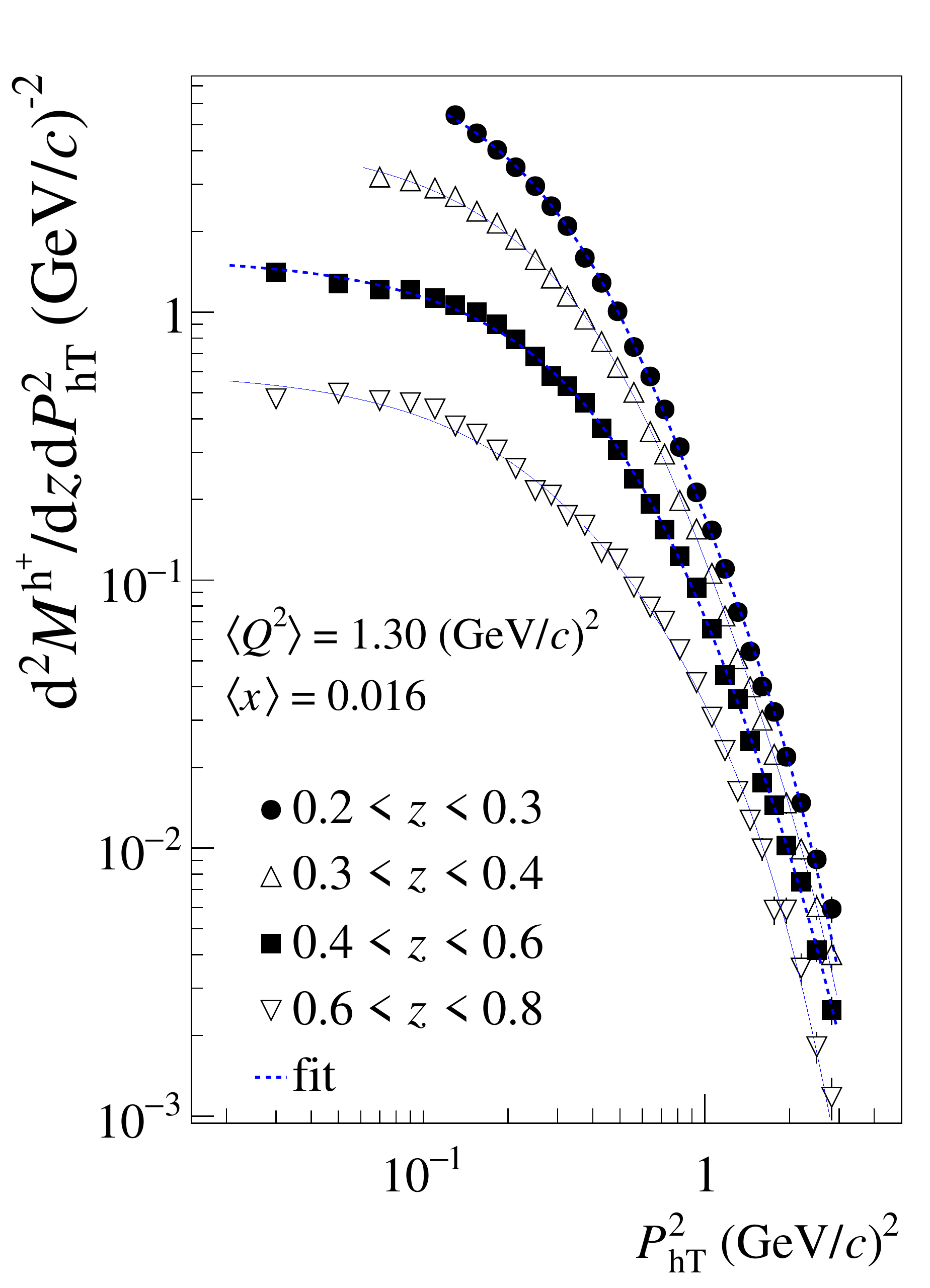}}
\caption{(a) Multiplicities of positively charged hadrons as a function of $P_{\rm{hT}}^{2}$ in four $z$ bins at $\langle Q^2 \rangle=1.25$ (GeV/$c$)$^{2}$ for $\langle x\rangle=0.006$. The curves correspond to the fits using the sum of two exponentials (Eq.~\ref{2expeq}). Only statistical uncertainties are shown and used in the fit. (b) Same as (a) for $\langle x\rangle=0.016$.}
\label{muland2exp}
\end{figure}

\begin{figure}[!h]
\centering
\includegraphics[height=10.5cm,width=.75\textwidth]{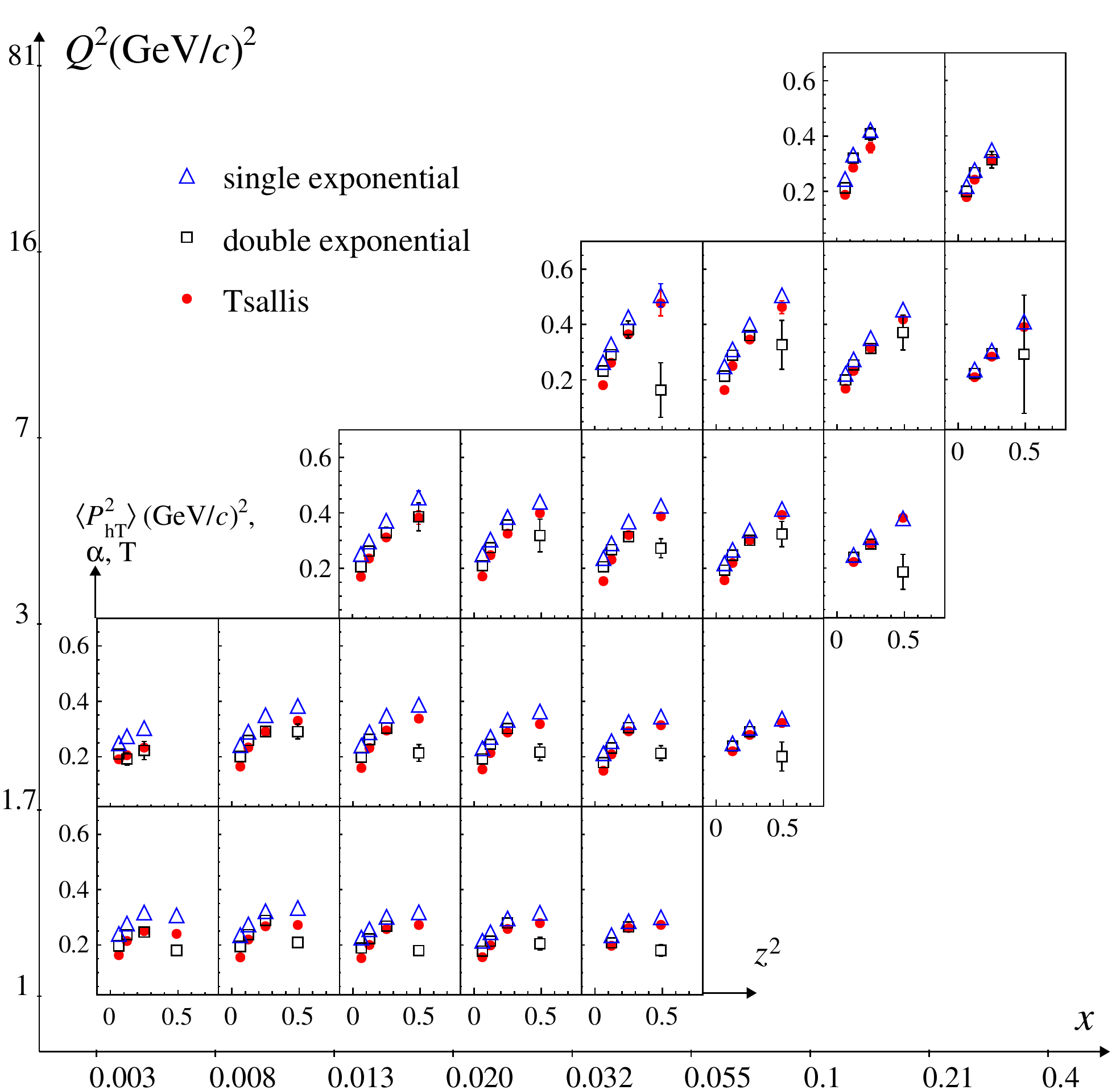}
\caption{Average transverse momentum obtained from the fit of h$^{+}$ multiplicities using the three fit functions given in Eqs.~\ref{1expfunction},~\ref{2expeq},~\ref{tsalliseq}: $\langle P_{\rm{hT}}^2\rangle$, $\alpha_1$ and $T$ as a function of $z^2$ in ($x$,~$Q^2$) bins.}
\label{avg12exptsallis}
\end{figure}

The first function ($F_{1}$) is defined as the sum of two single-exponential functions (Eq.~\ref{2expeq}). While $N_1$ and $N_1^{\ensuremath{\prime}}$ denote the normalisation coefficients, $\alpha_1$ and $\alpha_1^{\ensuremath{\prime}}$ denote the inverse slope coefficients of the first and the second exponential function, respectively. All coefficients depend on $x$, $Q^2$ and $z$. Figure~\ref{muland12exp} shows in a typical ($x$,~$Q^2$,~$z$) bin the multiplicities of positively charged hadrons as a function of $P_{\rm{hT}}^{2}$ fitted using $F_{1}$. As described above for Ref.~\cite{Anselmino:2006rv}, the two exponential functions in our parameterisation $F_{1}$ can be attributed to two completely different underlying physics mechanisms that overlap in the region $P_{\rm{hT}}^{2} \simeq 1$~(GeV/$c$)$^{2}$.
%The first function is attributed to leading order virtual photon absorption process, where the hadron transverse momentum $P_{\rm{hT}}$ is due only to the intrinsic transverse momentum $k_\perp$ of quarks in the nucleon and to the fragmentation process. The second function accounts for contributions of higher order pQCD processes to the $P_{\rm{hT}}$ spectrum.
% The two contributions match in the overlapping region, i.e. $P_{\rm{hT}}^{2} \sim 1$~(GeV/$c$)$^{2}$, as shown by the red curve.
%
Figure~\ref{muland2exp} shows, as an example, multiplicities of positively charged hadrons as a function of $P_{\rm{hT}}^{2}$, measured at $\langle Q^2\rangle \sim 1.25$~(GeV/$c$)$^2$ for two bins of $x$ with average values $\langle x\rangle=0.006$ and $\langle x\rangle=0.016$, in the four $z$ bins. Only statistical uncertainties are shown and used in the fit. Values of $\chi^{2}_{\rm{dof}}$ of about 1 are obtained in all ($x$,~$Q^2$,~$z$) bins, except for a few (6 out of 81) bins, where where values as small as 0.52 and as large as 2.52 are obtained.
The normalisation coefficients $N_{1}$ and $N_{1}^{\ensuremath{\prime}}$ are found to have a strong variation with $x$ and $z$ and a rather weak variation with $Q^2$, reflecting the ($x$,~$Q^2$) dependence of collinear PDFs and the $z$-dependence of collinear FFs. The inverse slope $\alpha_{1}$ has an average value of about 0.23~(GeV/$c$)$^2$ for $Q^{2} < 3$~(GeV/$c$)$^2$ and about 0.28~(GeV/$c$)$^2$ for larger values of $Q^2$. Its dependence on $z^2$ is discussed below using Fig.~\ref{avg12exptsallis}.
The inverse slope $\alpha_1^{\ensuremath{\prime}}$ has an average value of about 0.6~(GeV/$c$)$^2$ and shows a rather weak variation with $x$ and $Q^2$.

%~~~~~~~~~~~~~~~~~~~~~~~~~~~~~~~~~~~~~~~~~~~~%
%                                                                                            %
%		FIT USING TSALLIS FUNCTION	                   %
%~~~~~~~~~~~~~~~~~~~~~~~~~~~~~~~~~~~~~~~~~~~~%

The so-called Tsallis function $F_{2}$~\cite{Wilk:2008ue}, see Eq.~\ref{tsalliseq}, describes the two different kinds of power-law behaviour in the two regions of $P_{\rm{hT}}$ through a single function. The advantage of this function is that it provides both the inverse slope parameter $T$ that characterises the small-$P_{\rm{hT}}$ range and the exponent $1/(1-q)$ that parameterises the power-law tail at large $P_{\rm{hT}}$. The charged-hadron multiplicities (Figs.~\ref{fullmul0}-\ref{fullmul3}) are fitted in each ($x$,~$Q^2$,~$z$) bin using only statistical uncertainties. Reasonable values of $\chi^{2}_{\rm{dof}}$ are obtained in most bins except for 11 out of 81 bins where they are larger than 2 reaching up to $3.65$.
The exponent parameter $q$ has an average value of about 1.2. The exponent $1/(1-q)$ strongly depends on $x$ with a weaker dependence on $z$ and no variation with $Q^2$, while its $z$-dependence is observed to increase with $x$.
The inverse slope parameter $T$ ranges between 0.15~(GeV/$c$)$^2$ and 0.4~(GeV/$c$)$^2$ and shows a significant non-linear dependence on $z^{2}$ over the full $z$-range.

%~~~~~~~~~~~~~~~~~~~~~~~~~~~~~~~~~~~~~~~~~~~~%
%                                                                                            %
% Comparison for the inverse slope between all 3 functions %
%~~~~~~~~~~~~~~~~~~~~~~~~~~~~~~~~~~~~~~~~~~~~%

The inverse slopes $\langle P_{\rm{hT}}^{2}\rangle$, $\alpha_1$ and $T$, which were obtained using the fitting functions given in equations~\ref{1expfunction},~\ref{2expeq} and~\ref{tsalliseq} respectively, are presented and compared to each other as a function of $z^2$ in $(x,~Q^2)$ bins in Fig.~\ref{avg12exptsallis}. A weak non-linear dependence on $z^2$ is observed at small $x$ and $Q^2$, which becomes more pronounced at larger values of $Q^2$. The inverse slope T reproduces the same $z^2$-dependence as that of $\langle P_{\rm{hT}}^2\rangle$ described in Fig.~\ref{1expfunction}. It is observed to be in fair agreement with $\alpha_1$ except for $z > 0.6$.

%~~~~~~~~~~~~~~~~~~~~~~~~~~~~~~~~~~~~~~~~~~~~%
%                                                                                            %
%		Comparison data versus fit
%~~~~~~~~~~~~~~~~~~~~~~~~~~~~~~~~~~~~~~~~~~~~%

\begin{figure}[!h]
\centering
\subfigure[]{\label{a}\includegraphics[height=9.cm,width=.4\textwidth]{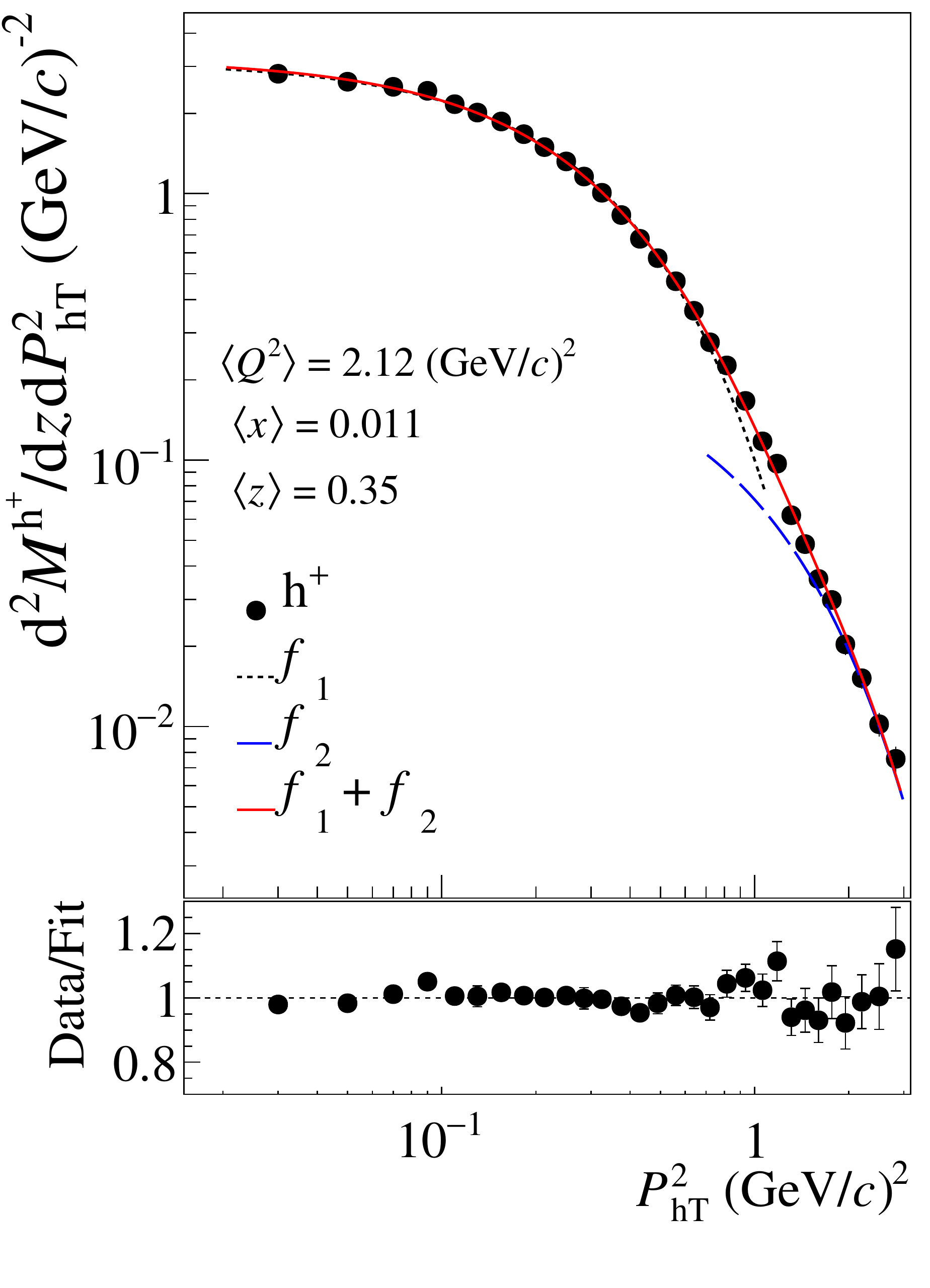}}
\subfigure[]{\label{c}\includegraphics[height=9.cm,width=.4\textwidth]{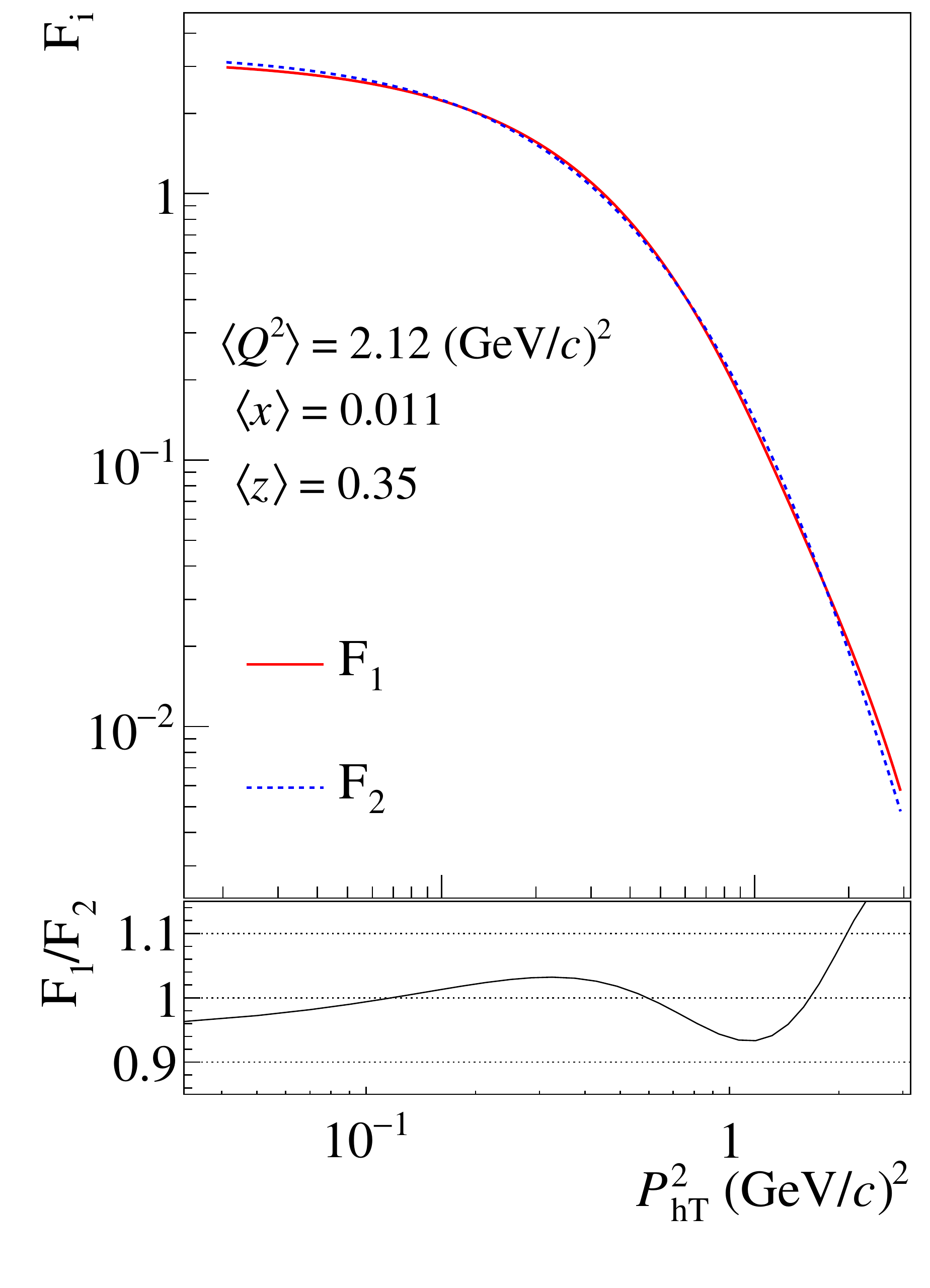}}
\caption{(a) Upper panel: Multiplicities of positively charged hadron as a function of $P_{\rm{hT}}^{2}$ for $\langle Q^2 \rangle=2.12$~(GeV/$c$)$^{2}$, $\langle x\rangle=0.011$ and $\langle z\rangle=0.35$. The black dotted curve represents the first exponential function of Eq.~\ref{2expeq}, the blue dashed curved represents the second exponential function of Eq.~\ref{2expeq}, and the red curve represents the sum. 
Only statistical uncertainties are shown and used in the fit. Lower panel: The ratio of the experimental points to the fit as a function of $P_{\rm{hT}}^2$. (b) Comparison between the fits obtained using F$_1$ (Eq.~\ref{2expeq}) and F$_2$ (Eq.~\ref{tsalliseq}) in the same kinematic bin as in (a).}
\label{comparisondatafit}
\end{figure}

A comparison between the data and the fit function $F_{1}$ is shown in Fig.~\ref{a} in a typical kinematic bin with  $\langle Q^{2}\rangle=2.12$~(GeV/$c$)$^2$ and $\langle x\rangle = 0.011$. The upper panel shows the multiplicities of positive hadrons as a function of $P_{\rm{hT}}^{2}$ and the corresponding fit function, and the lower panel shows the ratio between the data and the fit. A comparison between the two fitting functions $F_{1}$ and $F_{2}$ is shown in Fig.~\ref{c} for the same ($x$,~$Q^2$,~$z$) bin.
The $P_{\rm{hT}}^{2}$-dependence of $h^{+}$ multiplicities is equally well described by the two functions $F_{1}$ and $F_{2}$, as can be seen from the ratio in Fig.~\ref{c}. The same agreement is obtained for negatively charged hadrons.

%~~~~~~~~~~~~~~~~~~~~~~~~~~~~~~~~~~~~~~~~~~~~~~%
%     Multiplicities fitted using 1 and 2 exponentials fit curves     %

%~~~~~~~~~~~~~~~~~~~~~~~~~~~~~~~~~~~~~~~~~~~~~~%
%	Multiplicities fitted using TWO EXPONENTIALS                %

%~~~~~~~~~~~~~~~~~~~~~~~~~~~~~~~~~~~~~~~~~~~~~~%

%~~~~~~~~~~~~~~~~~~~~~~~~~~~~~~~~~~~~~~~~~~~~~~%
%                    COMPARISON DATA versus FIT

%-----Summary-----
\clearpage 
\section{\large{Summary}}
\label{sumary}
\vspace{0.3cm}

We have measured differential multiplicities of charge-separated hadrons in semi-inclusive measurements using muons of 160 GeV/$c$ impinging on an isoscalar (deuteron) target. Using a high-statistics data set collected in 2006, the measurement covers a wide kinematic domain of $Q^{2} > 1$~(GeV/$c$)$^2$, $W > 5$~GeV/$c^2$, $0.003 < x < 0.4$, $0.2< z <0.8$ and 0.02~(GeV/$c$)$^2$ $< P_{\rm{hT}}^2 <$ 3~(GeV/$c$)$^2$. The results are presented as a function of the square of the hadron transverse momentum $P_{\rm{hT}}^{2}$ in three-dimensional bins of $x$, $Q^2$ and $z$, which leads to a total of 4918 experimental data points. The numerical values are available on HepData~\cite{hepDATA} with and without subtraction of the estimated contribution of diffractive vector-meson production in SIDIS.

The h$^+$ multiplicities are only slightly larger than the h$^-$ ones in most of the bins, while for large $x$ and $z$ this difference increases. No significant difference between h$^+$ and h$^-$ is observed in the shape of the $P_{\rm{hT}}^{2}$-dependence of the multiplicity. Both h$^+$ and h$^-$  multiplicities are observed to flatten  at very small values of $P_{\rm{hT}}^{2}$ in the kinematic region of low $x$ and $Q^2$ and large $z$, where contributions from diffractive vector-meson production are the highest. Our results are compared to earlier measurements of hadron multiplicities and  cross  sections  by EMC,  HERMES  and  JLab. Good  agreement  was  found  with EMC for $W^2<150$~(GeV/$c$)$^2$, although the EMC data were collected at different beam energies and with different targets. In order to compare with HERMES, we have integrated our multiplicities over the phase space that is common to both experiments. While  reasonable agreement is obtained at small $z$ and $P_{\rm{hT}}^{2}$, differences are observed for large $z$ and $P_{\rm{hT}}$ where neither magnitudes nor $P_{\rm{hT}}^{2}$-dependences agree.
The $\pi^+$ semi-inclusive cross section measured by the E00-18 experiment at JLab shows fair agreement with COMPASS h$^+$ multiplicities, albeit with some discrepancy in the $P_{\rm{hT}}$-dependence that might be explained by the difference in the kinematic ranges of the measurements.

In the range of small-$P_{\rm{hT}}^{2}$, i.e. $P_{\rm{hT}}^{2} < 1$~(GeV/$c)^{2}$, the measured multiplicities were successfully fitted using a single Gaussian parameterisation. A non-linear $z^2$-dependence of the average transverse momentum is observed in the range of small $x$ and $Q^2$, which confirms the conclusions of Ref.~\cite{Adolph:2013stb}, while it is almost linear for large values of $x$ and $Q^2$. In order to fit the multiplicities over the full $P_{\rm{hT}}$-range measured by COMPASS, a more complex functional form is required, i.e. either a sum of two Gaussian functions or the so-called Tsallis function. All fits reproduce the data well and their inverse slopes agree well with one another already when using only statistical uncertainties in the fits.

\section{\large{Acknowledgements}}
\vspace{0.3cm} 

We gratefully acknowledge the support of the CERN management and staff and the skill and effort of the technicians of our collaborating institutes. This work was made possible by the financial support of our funding agencies.

\clearpage
\bibliographystyle{prsty}
%\bibliography{Chargedhadrons.bib}

\end{document}